\documentclass[aps,pra,twocolumn,showpacs,preprintnumbers,amsmath,amssymb,superscriptaddress]{revtex4-1}

\usepackage{graphicx}
\usepackage{dcolumn}
\usepackage{bm}
\usepackage{epsfig}
\usepackage{amsmath}
\usepackage{amssymb}
\usepackage{color}
\usepackage{bbm}
\usepackage{braket}
\usepackage[colorlinks=true,citecolor=blue,linkcolor=blue,urlcolor=blue]{hyperref}
\usepackage{tikz}

\usepackage{adjustbox}
\newcommand{\im}{\operatorname{i}\!}
\newcommand{\ex}{\operatorname{e}\!}
\newcommand{\id}{\operatorname{d}\!}

\begin{document}

\title{Quantum computing with superconducting circuits in the picosecond regime}

\author{Daoquan Zhu}
\affiliation{State Key Laboratory for Mesoscopic Physics, School of Physics, Frontiers Science Center for Nano-optoelectronics $\&$ Collaborative Innovation Center of Quantum Matter, Peking University, Beijing 100871, China}
\affiliation{Collaborative Innovation Center of Extreme Optics, Shanxi University, Taiyuan, Shanxi 030006, China}
\author{Tuomas Jaako}
\affiliation{Vienna Center for Quantum Science and Technology, Atominstitut, TU Wien, 1040 Vienna, Austria}

\author{Qiongyi He}
\email{qiongyihe@pku.edu.cn}
\affiliation{State Key Laboratory for Mesoscopic Physics, School of Physics, Frontiers Science Center for Nano-optoelectronics $\&$ Collaborative Innovation Center of Quantum Matter, Peking University, Beijing 100871, China}
\affiliation{Collaborative Innovation Center of Extreme Optics, Shanxi University, Taiyuan, Shanxi 030006, China}

\author{Peter Rabl}
\email{peter.rabl@tuwien.ac.at}
\affiliation{Vienna Center for Quantum Science and Technology, Atominstitut, TU Wien, 1040 Vienna, Austria}

\date{\today}

\begin{abstract}
We discuss the realization of a universal set of ultrafast single- and two-qubit operations with superconducting quantum circuits and investigate the most relevant physical and technical limitations that arise when pushing for faster and faster gates. With the help of numerical optimization techniques, we establish a fundamental bound on the minimal gate time, which is determined independently of the qubit design solely by its nonlinearity. In addition, important practical restrictions arise from the finite qubit transition frequency and the limited bandwidth of the control pulses. We show that for highly anharmonic flux qubits and commercially available control electronics, elementary single- and two-qubit operations can be implemented in about 100 picoseconds with residual gate errors below $10^{-4}$. Under the same conditions, we simulate the complete execution of a compressed version of Shor's algorithm for factoring the number 15 in about one nanosecond. These results demonstrate that compared to state-of-the-art implementations with transmon qubits, a hundredfold increase in the speed of gate operations with superconducting circuits is still feasible.
\end{abstract}

\maketitle

\section{Introduction}
Current efforts to build large-scale quantum computers are motivated by the prospect of enabling quantum algorithms with an exponential or at least a quadratic speedup for solving certain types of hard computational problems~\cite{NielsenChuang}. This speedup implies that independently of the time it takes to execute a single gate, a quantum computer will outperform its classical counterpart when the size of the problem is sufficiently large. Also, in first physical realizations of small quantum processors it has turned out that it is usually beneficial to encode qubits in weakly interacting degrees of freedom, for example, using spin states instead of electronic orbitals. These qubits are typically slower in their control, but also exhibit much longer decoherence times such that overall more coherent gate operations can be performed. Thus, for both conceptual and experimental reasons, the search for fast quantum gates has so far only played a secondary role in the development of quantum technologies. However, for any real-world application not only the scaling, but also the total computation time will be of importance. In addition, in every physical device  there will be decoherence processes that are very hard or even impossible to avoid. In this case the realization of faster gate operations becomes a necessity to further improve the fidelity of the computation.

In the development of superconducting quantum computers~\cite{Krantz2019}, many important breakthroughs were facilitated by the development of the transmon~\cite{Koch2007} and related qubit designs, where quantum information is encoded in the two lowest levels of a weakly anharmonic oscillator. Typical operation timescales for such transmon qubits are in the order of a few tens to a few hundreds of nanoseconds, but with the help of optimized control pulses~\cite{Spoerl2007,Montangero2007,Motzoi2009,Rebentrost2009,Safei2009,Lucero2010,Gambetta2010,Egger2014,Schutjens2013,Motzoi2013,Huang2014,Theis2016,Liebermann2016,Kirchhoff2018,Theis2018,Machnes2018,GarciaRipoll2020,Xu2020,Watts2015,Goerz2015b,Glaser2015,Leung2017,Goerz2017,Abdelhafez2020,Shillito2020,Abdelhafez2019,Tian2020} the implementation of single- (two-) qubit gates in about $ 4\,\mathrm{ns} $ \cite{Chow2010,Werninghaus2020} ($ 12\,\mathrm{ns} $ \cite{Arute2019,Foxen2020}) has been demonstrated. To realize much faster operations it is necessary to use qubits with higher nonlinearities, such as flux qubits~\cite{Yan2016}. In this case, single-qubit rotations with a duration of only $ 1.6\,\mathrm{ns} $ and residual errors in the range of $10^{-3}$ have  been recently achieved \cite{Yurtalan2020}. Even stronger nonlinearities can be reached with charge qubits, for which single- and two-qubit gates with a duration of $ O(100\,\mathrm{ps})$ have been implemented already at the very early stage of this field~\cite{Yamamoto2003}. However, because of their rather shorter coherence times and other difficulties in their control, charge qubits are currently not considered as a suitable candidate for high-fidelity quantum computation. On the theoretical side, single- and two-qubit gates with durations of $ 0.1\,\mathrm{ns} $ and below have been predicted~\cite{Spoerl2007,Montangero2007,Motzoi2009,Romero2012,Huang2014}, but either for charge qubits or based on crucial approximations, such as two- or three-level truncations or instantaneously switchable control fields. Therefore, despite its long-term relevance, the 
implementation of quantum gates in the picosecond regime is still little explored and the intriguing question about the ultimate limit for the speed of superconducting quantum processors remains open.

In this paper we present a systematic numerical study on the implementation of ultrafast single- and two-qubit gates in superconducting circuits. For this analysis we consider a generic circuit design with a varying degree of nonlinearity, which interpolates between the most common types of qubits in use today. By applying optimal control techniques we determine the maximal fidelities for a set of elementary single- and two-qubit gates that can be achieved for a given gate time $t_g$. In these numerical simulations we take the full multilevel structure of the underlying circuit into account and include as well  the effects of a finite bandwidth of the control pulses. In this way we obtain a realistic relation between the minimal achievable gate time, $t_g^{\rm min}$, and the most relevant circuit and control parameters. In the limit of control pulses with infinite bandwidth we identify a fundamental bound $t_g^{\rm min}\gtrsim C/|\alpha|$, where $\alpha$ is the circuit nonlinearity and $C=2\pi$ ($C=\pi$) for single- (two-) qubit gates. This simple scaling holds over several orders of magnitude and can be used to optimize the qubit design in cases where it is not reached. For circuits with a limited controllability, important practical restrictions arise both from the finite bandwidth of the control pulses as well as from a finite qubit oscillation period. These limitations become highly relevant for gate times around $t_g\sim 100$ ps and below, but can in principle be overcome by using more advanced control electronics and implementing multi-axis control schemes.  

 In a second step we go beyond individual quantum gates and discuss the  implementation of composite quantum circuits consisting of multiple ultrafast gates applied in a consecutive manner.  Here we encounter additional clocking requirements, which are not yet of relevance  at current operation speeds. As an illustrative example, we consider a minimal, but still useful circuit, which implements a compressed version of Shor's algorithm for factoring the number 15~\cite{Monz2016}.  A full numerical simulation of the whole gate sequences shows that this quantum algorithm can be executed in about one nanosecond with an overall fidelity of $\mathcal{F}\approx 0.999$, assuming realistic qubit parameters and state-of-the-art waveform generators. These results demonstrate the principle feasibility of superconducting quantum circuits operated in the picosecond domain and provide a valuable benchmark for further refined theoretical and experimental studies in this direction.

\section{A superconducting quantum processor}\label{sec:model}
For the following analysis we consider a linear quantum processor, which consists of an array of $N$ identical superconducting qubits with individual local control and switchable nearest-neighbor interactions. The Hamiltonian for the whole processor can be written as
\begin{equation}
H(t)= \sum_{i=1}^N H_q^{(i)}(t) + \sum_{i=1}^{N-1} H_{qq}^{(i,i+1)}(t),
\end{equation} 
where $H_q(t)$ and $H_{qq}(t)$ represent the individual qubit Hamiltonian and the qubit-qubit interaction, respectively.

\begin{figure}[t]
    \begin{center}
	    \includegraphics[width=\columnwidth]{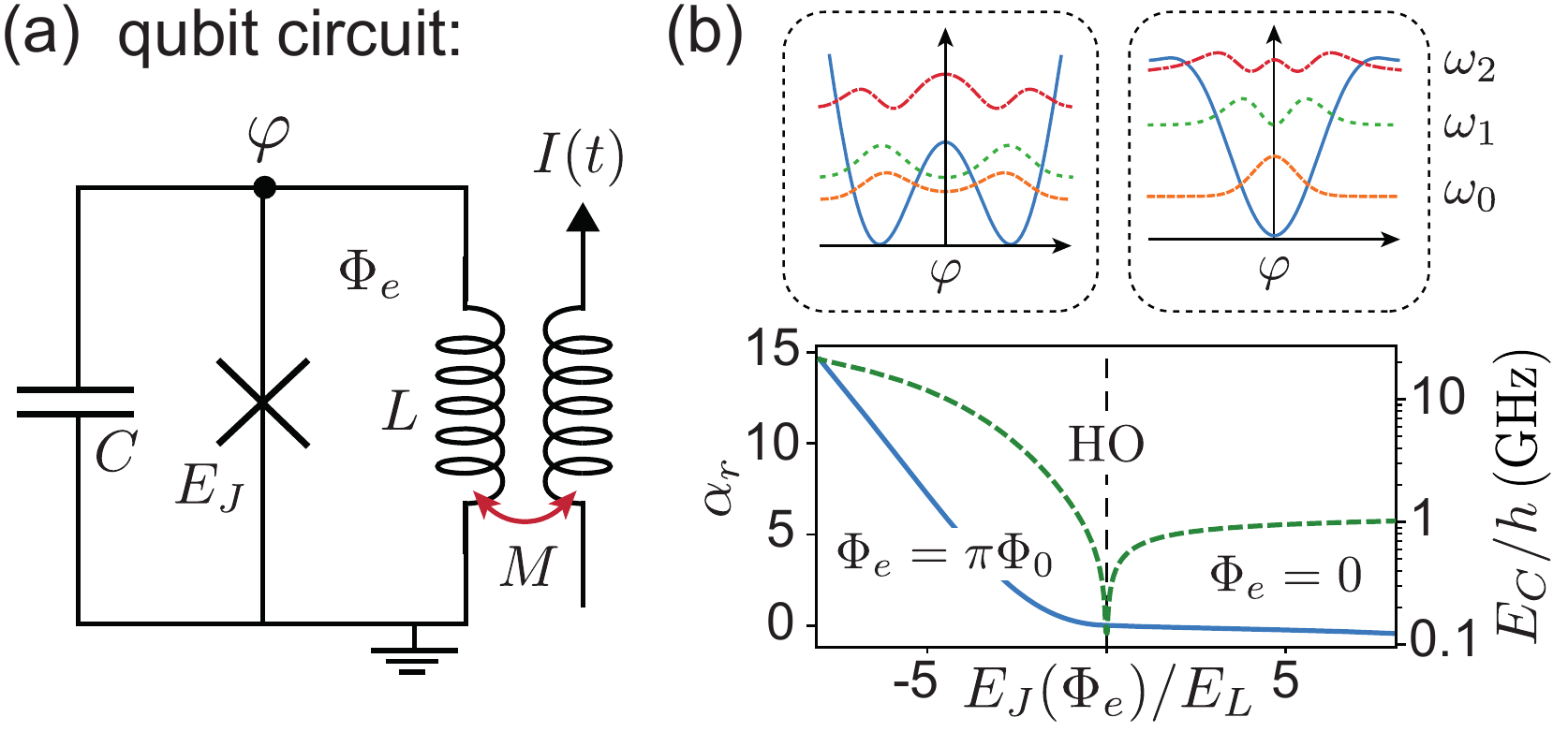}
	    \caption{(a) Circuit for a generic superconducting qubit formed by a capacitor $C$, an inductor $L$ and a Josephson junction with Josephson energy $E_J$. The qubit can be controlled by an external current $I(t)$, which couples to the dimensionless flux variable $\varphi$ through a mutual inductance $M$. (b) Plot of the nonlinearity parameter $\alpha_r$ (solid line) for different ratios of $E_J(\Phi_e)/E_L = E_J\cos(\Phi_e/\Phi_0)/E_L$ and $E_J/E_C=4$. The dashed line shows the corresponding value of $E_C$, which is used to obtain a qubit frequency of $\omega_{10}/(2\pi) = 5\,\mathrm{GHz}$. The two upper panels show a sketch of the potential for the phase variable $\varphi$ and the lowest three eigenstates for the two cases of $\Phi_e=\pi\Phi_0$ (left) and $\Phi_e=0$ (right).}
        \label{fig:1_Qubit}
    \end{center}
\end{figure}

\subsection{Superconducting qubits}
We assume that each qubit is implemented by the circuit shown in Fig.~\ref{fig:1_Qubit}(a), which consists of a capacitance $ C $, an inductance $ L $ and a Josephson junction with Josephson energy $ E_J $ connected in parallel. The superconducting loop formed by $ L $ and $ E_J $ is threaded by a static magnetic flux $ \Phi_e $. In addition, the qubit circuit can be driven by a time-dependent external current $I(t)$. We write the full qubit Hamiltonian as $H_q(t)=H_0+ H_{\rm drive}(t)$, where 
\begin{align}\label{eq:h0}
    H_0 = 4E_C q^2+ \dfrac{E_L}{2}\varphi^2  - E_J\cos\left( \varphi + \Phi_e/\Phi_0 \right).
\end{align}
Here $E_C=e^2/(2C)$, $E_L=\Phi_0^2/L$ and $ \Phi_0 = \hbar/(2e) $ is the reduced flux quantum. 
The dimensionless charge and flux operators $q$ and $\varphi$ obey the commutation relations $[\varphi,q]=\im$ . The external driving term is given by
\begin{align}
    H_{\rm drive}(t) = \dfrac{MI(t)\Phi_0}{L} \varphi,
\end{align}
where $M$ is the mutual inductance.

Depending on the values of $E_C$, $E_L$, $E_J$ and $\Phi_e$, this basic circuit can be operated in different regimes and be used to realize qubits with different degrees of nonlinearity, as sketched in Fig.~\ref{fig:1_Qubit}(b). Taking the external flux to be $ \Phi_e =0 $ the potential landscape for the phase coordinate $\varphi$ has a single minimum and for $(E_L+E_J)>E_C$ we obtain a transmon-type qubit with a weakly negative anharmonicity. Instead, a value of $\Phi_e = \pi\Phi_0 $ and $E_L<E_J$ results in a double-well potential with a positive anharmonicity, which is representative for various types of flux qubits. In both cases the nonlinearity can be tuned by adjusting the charging energy $E_C$. In the following we use this flexibility to investigate superconducting qubits with varying (relative) nonlinearity parameter
\begin{equation}
\alpha_{r}= \frac{E_2-E_1}{E_1-E_0}-1,
\end{equation}
where $E_n=\hbar \omega_n$ is the energy of the $n$-th eigenstate $|\psi_n\rangle$ of $H_0$. To do so we assume a fixed ratio of $E_J/E_C$ and vary  $E_L/E_C$, setting either $ \Phi_e =0 $ or $ \Phi_e =\pi\Phi_0 $. At the same time we adjust the absolute value of $E_C$ to keep the transition frequency between the lowest two eigenstates, $\omega_{10}=(E_1-E_0)/\hbar$, fixed. In Fig.~\ref{fig:1_Qubit}(b) we plot the resulting nonlinearity parameter for a ratio of $E_J/E_C=4$ and a characteristic transition frequency of $\omega_{10}/(2\pi)=5\,\mathrm{GHz}$.

For a given set of circuit parameters the bare Hamiltonian $H_0$ can be diagonalized and written as
\begin{equation}
H_0= \sum_n \hbar \omega_n |\psi_n\rangle\langle \psi_n|.
\end{equation}
To identify the actual qubit states $|0\rangle$ and $|1\rangle$ for encoding quantum information we change to an interaction picture with respect to $H_0$ and set
\begin{equation}\label{eq:Qubits}
|0\rangle = \ex^{-\im\omega_0 t}|\psi_0\rangle, \qquad  |1\rangle = \ex^{-\im\omega_1 t}|\psi_1\rangle.
\end{equation}
According to this definition a superposition of these qubit states does not evolve in time when $H_{\rm drive}=0$. In the interaction picture the effect of the driving term is given by 
\begin{equation}\label{eq:h_drive}
\tilde H_{\rm drive}(t) =\hbar \Omega(t)\sum_{n,m} \langle \psi_n|\varphi |\psi_m\rangle \ex^{\im\,(\omega_n-\omega_m)t}  |\psi_n\rangle \langle \psi_m|,
\end{equation}
where $\Omega(t)= MI(t)\Phi_0/(\hbar L)$ is the characteristic driving strength. In the limit of a weak resonant driving signal, $\Omega(t)=\Omega\cos(\omega_{10} t+\phi_d)$ and  $\Omega\ll \omega_{10}$, the control Hamiltonian reduces to  
\begin{equation}\label{eq:WeakDriving}
\tilde H_{\rm drive}(t)\simeq  \frac{\hbar \Omega \varphi_{10} }{2}\left(\ex^{\im\phi_d} |0\rangle\langle 1| +  \ex^{-\im\phi_d} |1\rangle\langle 0|\right),
\end{equation}
where $\varphi_{10}=\langle 1|\varphi|0\rangle$. In this limit, $\tilde H_{\rm drive}(t)$ allows us to implement qubit rotations along any axis in the $x,y$ plane by adjusting the phase $\phi_d$. However, it is important to keep in mind that on timescales comparable to $\omega^{-1}_{10}$, $\tilde H_{\rm drive}(t)$ represents a `one-axis control', since it only couples to the phase variable $\varphi$ and not to the conjugate charge.

\subsection{Qubit-qubit interactions}

\begin{figure}[t]
    \begin{center}
        \includegraphics[width=\columnwidth]{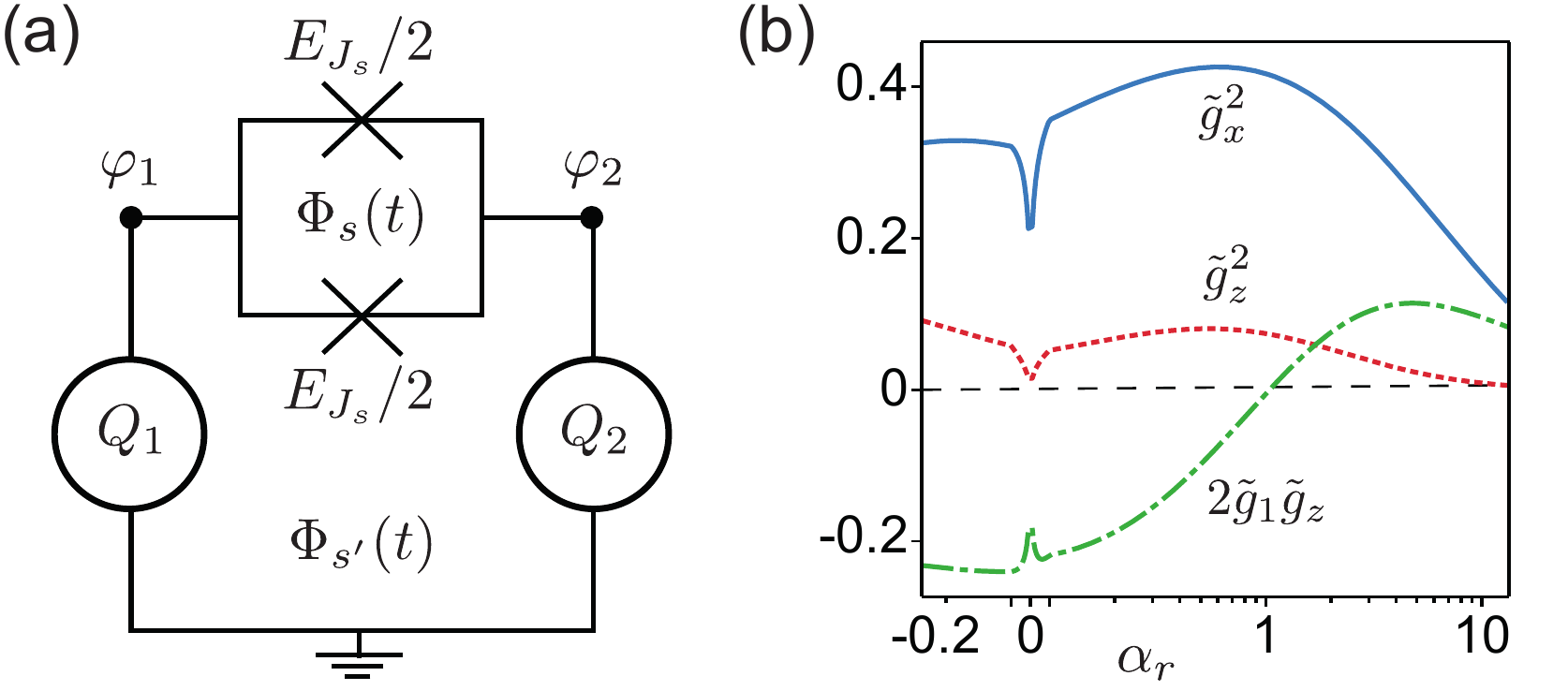}
    \end{center}
    \caption{Qubit-qubit interactions. (a) The qubits of Fig.~\ref{fig:1_Qubit}(a), represented by the circles, are coupled via a SQUID loop. The SQUID loop is formed by two Josephson junctions with the same Josephson energy $ E_{J_s}/2 $ and it is threaded by an external flux $ \Phi_s $. The external flux through the big loop formed by the lower arm of the SQUID and the qubits is denoted by $ \Phi_{s'} $. (b) Plot of the dimensionless coupling parameters defined in Eq.~\eqref{eq:CouplingParameters} as a function of the nonlinearity parameter $ \alpha_r$. For this plot the same circuit parameters as in Fig.~\ref{fig:1_Qubit}(b) have been assumed. Note that in (b) the scale of the $x$ axis is linear in the range $ [-0.1, 0.1] $ and logarithmic elsewhere.}
\label{fig:2_Interactions}
\end{figure}

For the implementation of two-qubit gates we consider a circuit as shown in Fig.~\ref{fig:2_Interactions}(a), where two neigboring qubits are coupled via a superconducting quantum interference device (SQUID) loop. The SQUID loop is formed by two identical Josephson junctions with Josephson energies $ E_{J_s}/2 $ and the loop is threaded by an external flux $ \Phi_s $. Another closed loop is formed by the lower junction of the SQUID, the inductor of the qubit on the left and the Josephson junction of the qubit on the right. The flux through this loop is denoted by $ \Phi_{s'} $. At a flux sweet spot of the qubits, $ \Phi_e = n\pi\Phi_0 $, where $n$ is an integer, the Hamiltonian for this coupling element is
\begin{equation}
\begin{split}
    H_{qq} = \dfrac{E_{J_s}}{2}&  \left[ \cos\left( \Delta\varphi + \dfrac{ \Phi_{s'}}{\Phi_0} \right) \right.\\
   & + \left. \cos\left(\Delta\varphi + \dfrac{ \Phi_{s'} + \Phi_s}{\Phi_0} \right)\right],
\end{split}
\end{equation}
where $ \Delta\varphi = \varphi_1 - \varphi_2 $. By assuming time-dependent external fluxes with a fixed relation $\Phi_{s'} (t)= -\Phi_s(t)/2 $, the coupling simplifies to
\begin{align}
    H_{qq}(t) = E_{J_s}(t)\cos\left( \Delta\varphi \right),
\end{align}
with a tunable Josephson energy $E_{J_s}(t)=E_{J_s}\cos(\Phi_s(t)/2\Phi_0)$. By varying the value of $\Phi_s/\Phi_0$, the coupling can be tuned from 0 up to a maximal value of $E_{J_s}$. When restricted to the qubit subspace, we can write 
\begin{align}\label{eq:CouplingParameters}
    \cos\left(\varphi \right) = \tilde{g}_1\mathbbm{1} - \tilde{g}_z\sigma^z,\qquad \sin\left( \varphi  \right) = \tilde{g}_x\sigma^x,
\end{align}
where the $\sigma^k$ are the Pauli operators. Note that throughout this paper we adopt the convention $\sigma^z=|0\rangle\langle 0|-|1\rangle\langle 1|$, as usually assumed in the quantum computing literature. 
The dimensionless parameters $\tilde g_{1,z,x}$ depend on the specific qubit circuit, but in general all three of them are non-zero. Therefore, within this subspace the interaction Hamiltonian takes the form
\begin{align}
    H_{qq}(t)\simeq -\dfrac{\hbar\delta\omega(t)}{2}\left( \sigma_1^z + \sigma_2^z \right) + \hbar g_z(t)\sigma_1^z\sigma_2^z + \hbar g_x(t)\sigma_1^x\sigma_2^x ,
\end{align}
where $ \delta\omega(t) = 2\tilde{g}_1\tilde{g}_z E_{J_s}(t)/\hbar $, $ g_z(t)= \tilde{g}_z^2 E_{J_s}(t)/\hbar $ and $ g_x(t)= \tilde{g}_x^2 E_{J_s}(t)/\hbar $. In the transmon regime and for weak couplings, $H_{qq}(t)$ approximately reduces to a flip-flop interaction $\sim  g_x(t)(|01\rangle\langle 10| + |10\rangle\langle 01|) $, which can be used to realize a universal $\sqrt{\im {\rm SWAP}}$ gate~\cite{Krantz2019}. However, as shown in Fig.~\ref{fig:2_Interactions}(b), already for small nonlinearities also the $ZZ$ interaction and the single-qubit shifts are non-negligible and must be taken into account in the resulting two-qubit operation. 

For the actual numerical simulations of the gate we change again to the interaction picture, where the full coupling Hamiltonian reads
\begin{equation}\label{eq:h2q}
\begin{split}
\tilde H_{qq}(t)&=  E_{J_s}(t)\sum_{\substack{n_1,m_1\\n_2,m_2}}  \langle \psi_{n_1},\psi_{n_2}| \cos(\Delta \varphi) |\psi_{m_1},\psi_{m_2}\rangle \\
&\times  \ex^{\im\,(\omega_{n_1}+\omega_{n_2}-\omega_{m_1}-\omega_{m_2})t}  |\psi_{n_1},\psi_{n_2}\rangle \langle \psi_{m_1},\psi_{m_2}|.
\end{split}
\end{equation}
Since we assume that all qubits are identical and interact at most with one of their neighbors at a time, it is enough to analyze the evolutions generated by $\tilde H_{\rm drive}(t)$ and $\tilde H_{qq}(t)$ in order to model arbitrary quantum gates along the chain. 

Let us emphasize that although in our analysis we take the full multi-level dynamics of the single- and two-qubit circuits shown in Fig.~\ref{fig:1_Qubit}(a) and Fig.~\ref{fig:2_Interactions}(b) into account, these circuits and their control are still based on various idealizations. For example, we neglect the effect of any parasitic capacitive or inductive elements as well as any crosstalk between the control signals. In our simulations we will also neglect the effects of decoherence and decay, which assumes that large coherence times of $T^*_2\gtrsim 1\,\mathrm{\mu s}$~\cite{Yan2016,Kjaergaard2020} can be achieved independently of the circuit parameters and the degree of nonlinearity. Nevertheless, these imperfections do not directly affect the problem at hand, namely to identify the maximal speed of gate operations, and will thus not be considered in our analysis.

\section{Single-qubit gates}\label{sec:1q_gates}
We first discuss the implementation of single-qubit gates, which are generated through the local control Hamiltonian $\tilde H_{\rm drive}(t)$ in Eq.~\eqref{eq:h_drive}. The goal is to realize rotations of the form 
\begin{equation}
R_{k=X,Y,Z}(\theta)= \ex^{-\im\theta \sigma^k/2},
\end{equation}
within the qubit subspace spanned by the states $|0\rangle$ and $|1\rangle$. Due to limited control and transitions to other levels, these operations can only be implemented approximately in real circuits and we define by \cite{Motzoi2009,Theis2016}
\begin{equation}\label{eq:fidelity}
	\mathcal{F} = \dfrac{1}{d^2}\left| \mathrm{Tr}\left\{ U_{\rm target}^{\dagger}U(t_g) \right\} \right|^{2}
\end{equation}
the fidelity of the gate. Here $U_{\rm target}$ is the targeted unitary operation within the $d=2$ dimensional subspace and
\begin{equation}
U(t_g)= \mathcal{T} \ex^{-\im \int_0^{t_g} \id t \, \tilde H_{\rm drive}(t)},
\end{equation}
where $\mathcal{T}$ denotes the time-ordered exponential, is the actual evolution operator for the whole qubit circuit during the time interval $[0,t_g]$.

To maximize the fidelity $\mathcal{F}$ for a given total gate time $t_g$, we use coherent control techniques to find a numerically optimized shape for the control pulse $\Omega(t)=\Omega_c(t)$. As an ansatz for $\Omega_c(t)$ we use a pulse of the form  
\begin{equation}\label{eq:omega_c}
	\Omega_c(t)  = \Omega_0 \cos(\omega_{d} t+\phi_d) \sum_{n=1}^{n_{\rm max}} a_n\sin\left(\frac{ n\pi  t}{t_g}\right),
\end{equation}
where the overall strength of the driving field, $\Omega_0$, the frequency of the carrier, $ \omega_d $, the phase of the carrier, $\phi_d$, and the components of the pulse envelope, $a_n\in [-1,1]$, are adjustable parameters. We then use established numerical optimization algorithms to find the set of parameters that minimizes the gate error
\begin{equation}\label{eq:cost}
 \mathcal{E} =1 -\mathcal{F},
\end{equation}
using a fixed number of $n_{\rm max} = 20$ frequency components. While the resulting gate errors for individual data points might vary slightly, we find that none of the general trends and conclusions presented in this work depend significantly on the chosen ansatz for the pulse or the precise number of parameters, as long as $n_{\rm max}$ is sufficiently large. All details about the numerical procedure for determining $\Omega_c(t)$ are given in Appendix \ref{app:optimization}.

\subsection{The Hadamard gate}

\begin{figure}[t]
    \begin{center}
        \includegraphics[width=\columnwidth]{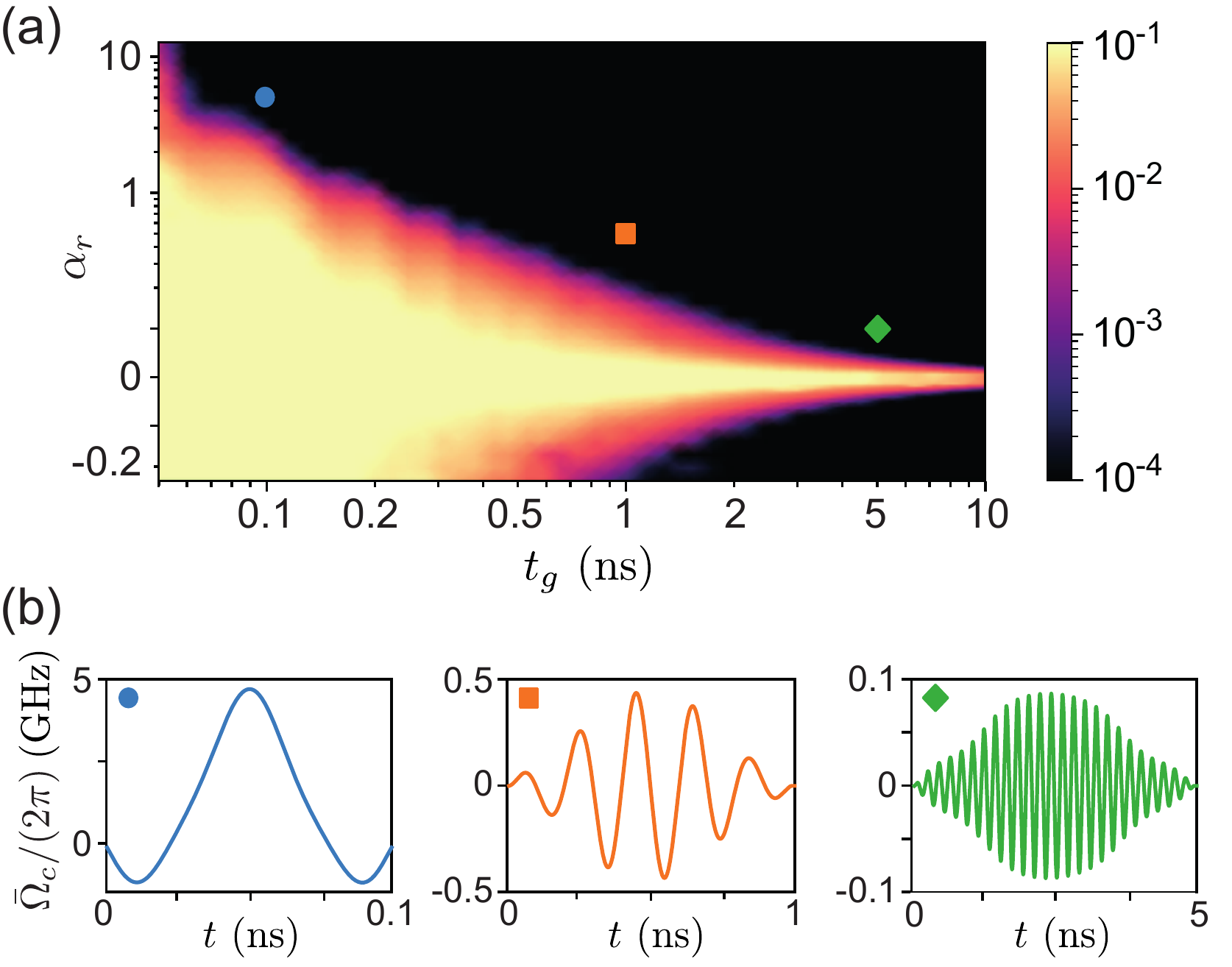}
        \caption{Optimization of the Hadamard-like gate, $ R_Y(\pi/2) $. (a) Plot of the numerically minimized gate error $\mathcal{E}=1-\mathcal{F}$ as a function of the gate time $t_g$  and the qubit nonlinearity parameter $\alpha_r$. (b) Examples of the optimized pulse shapes for $\bar \Omega_c(t)=\varphi_{10} \Omega_{c}(t)$ for the points marked in (a). Note that in (a) the scale of the $y$ axis is linear in the range $ [-0.1, 0.1] $ and logarithmic elsewhere.}
        \label{fig:3_H_Optimization}
    \end{center}
\end{figure}

Figure~\ref{fig:3_H_Optimization} summarizes the outcomes of this optimization for the example of a $\pi/2$ rotation about the $Y$-axis,
\begin{equation}
R_Y(\pi/2)= \frac{1}{\sqrt{2}}
\left(
\begin{array}{cc}
1& -1\\
1 & 1
\end{array}
\right),
\end{equation} 
which, up to a flip of the basis states, is just the usual Hadamard gate.
The main plot in Fig.~\ref{fig:3_H_Optimization} shows the minimal gate error $\mathcal{E}$ as a function of the gate time $t_g$ and the nonlinearity parameter of the qubit, $\alpha_r$. We see that for gate times down to about $t_g\approx 1\,\mathrm{ns}$, it is still possible to implement high-fidelity qubit rotations with errors $\mathcal{E}\lesssim 10^{-4}$ using qubits with moderate nonlinearities, $|\alpha_r|\simeq 0.25$. This degree of nonlinearity is still accessible with a transmon design, i.e., for qubits with a single potential minimum. The corresponding numerically optimized pulses shown in the center and right panel of Fig.~\ref{fig:3_H_Optimization}(b) simply consist of a near-resonant carrier tone with a slowly modulated amplitude. These findings are fully consistent with other optimal control studies \cite{Gambetta2010,Werninghaus2020,Safei2009,Goerz2017,Abdelhafez2019} and experiments \cite{Werninghaus2020,Chow2010,Lucero2010,Chow2012,Chow2009} with weakly nonlinear qubits.

For gate times below 1 ns the total duration of the $\pi/2$ pulse is already comparable to the bare rotation time of the qubit, $T_q=2\pi/\omega_{10}$, where $T_q=0.2\,\mathrm{ns}$ for the chosen qubit parameters. Nevertheless, for sufficiently high nonlinearities, it is still possible to implement rotations between the two lowest states of the circuit, with similar fidelities as above and in a time $t_g<T_q$. Although under these conditions the optimal control pulses are no longer very intuitive, they remain rather smooth and do not exhibit rapid oscillations or any other peculiar features.

\begin{figure}[t]
    \centering
    \includegraphics[width=\columnwidth]{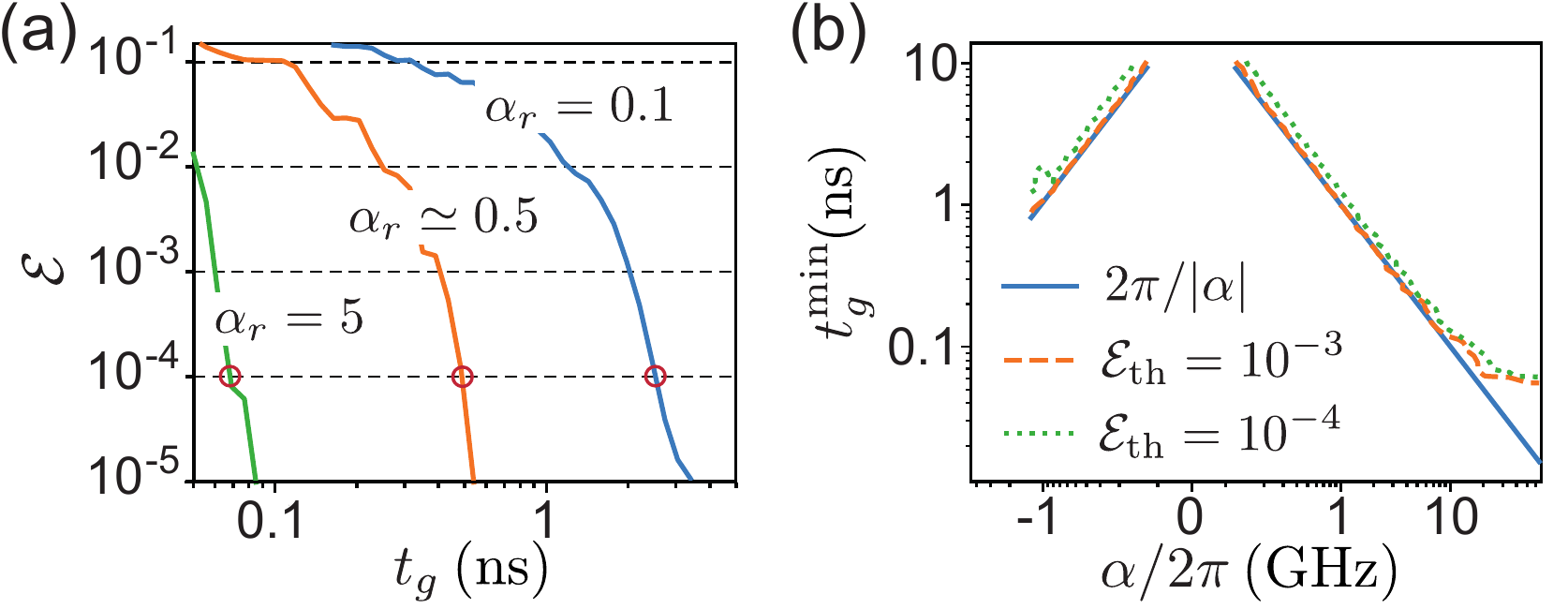}
    \caption{(a) Dependence of the gate error $\mathcal{E}$ for an $R_Y(\pi/2)$ rotation on the gate time $t_g$ and for three different values of the relative qubit nonlinearity. The red circles mark the times where the infidelity crosses the threshold of $\mathcal{E}_{\rm th}=10^{-4} $. (b) Minimal gate time $t_g^{\rm min}$ as a function of the absolute nonlinearity, $ \alpha=\omega_{10} \alpha_r $. The dashed and dotted lines show the results obtained for the error thresholds $ \mathcal{E}_{\rm th} = 10^{-3} $ and $ \mathcal{E}_{\rm th} = 10^{-4} $, respectively, whereas the solid line indicates the value of $t_g^{\rm min}=2\pi/|\alpha|$. Note that in (b) the scale of the $x$ axis is linear in the range $ [-0.1, 0.1] $ and logarithmic elsewhere.}
    \label{fig:4_Speed_Limit}
\end{figure}

In Fig.~\ref{fig:4_Speed_Limit}(a) we  show a cut through this infidelity map for different nonlinearity parameters. 
In the regime $10^{-5}<\mathcal{E}\leq 10^{-3}$ (note that our numerical search stops at a value of $\mathcal{E}= 10^{-5}$) the gate errors depend very sensitively on $t_g$, which allows us to identify for each $\alpha_r$ a minimal gate time $t^{\rm min}_g(\alpha_r)$, below which high-fidelity rotations are no longer possible. For concreteness we choose here a threshold value for the tolerable gate error of $\mathcal{E}_{\rm th}=10^{-4}$, which means $\mathcal{E}(t^{\rm min}_g)=\mathcal{E}_{\rm th}$. In Fig.~\ref{fig:4_Speed_Limit}(b) we plot this minimal gate time as a function of the absolute nonlinearity, $ \alpha = \alpha_r\omega_{10}$. These numerical results fit very well the analytic scaling 
\begin{equation}\label{eq:speed_limit}
t^{\rm min}_g(\alpha) \approx C \times \frac{2\pi}{|\alpha|},
\end{equation}
with a numerical constant $ C \approx 1.36 $ (for an error threshold of $\mathcal{E}_{\rm th}=10^{-3}$ the same fit yields $C\approx 1.06$). While for simple three-level models, where $\alpha$ is the only relevant energy scale, the scaling given in Eq.~\eqref{eq:speed_limit} is intuitively expected, we find that it holds surprisingly well for a large parameter range over which the level structure of the qubit circuit and all the coupling matrix elements vary considerably. For negative nonlinearities, this scaling breaks down for $|\alpha|/(2\pi)\gtrsim 1$ GHz due to accidental two-photon resonances with higher energy levels. For $\alpha>0$, we observe a deviation only for very large nonlinearities, which correspond to gate times of about $t_g\approx 50\,\mathrm{ps}$. Below we will provide an intuitive explanation for this behavior, which is a consequence of the finite qubit oscillation period, $T_q$.

\subsection{Arbitrary $R_{n(\phi)}(\theta)$ rotations}\label{subsec:RnRotations}

\begin{figure}[t]
    \centering
    \includegraphics[width=\linewidth]{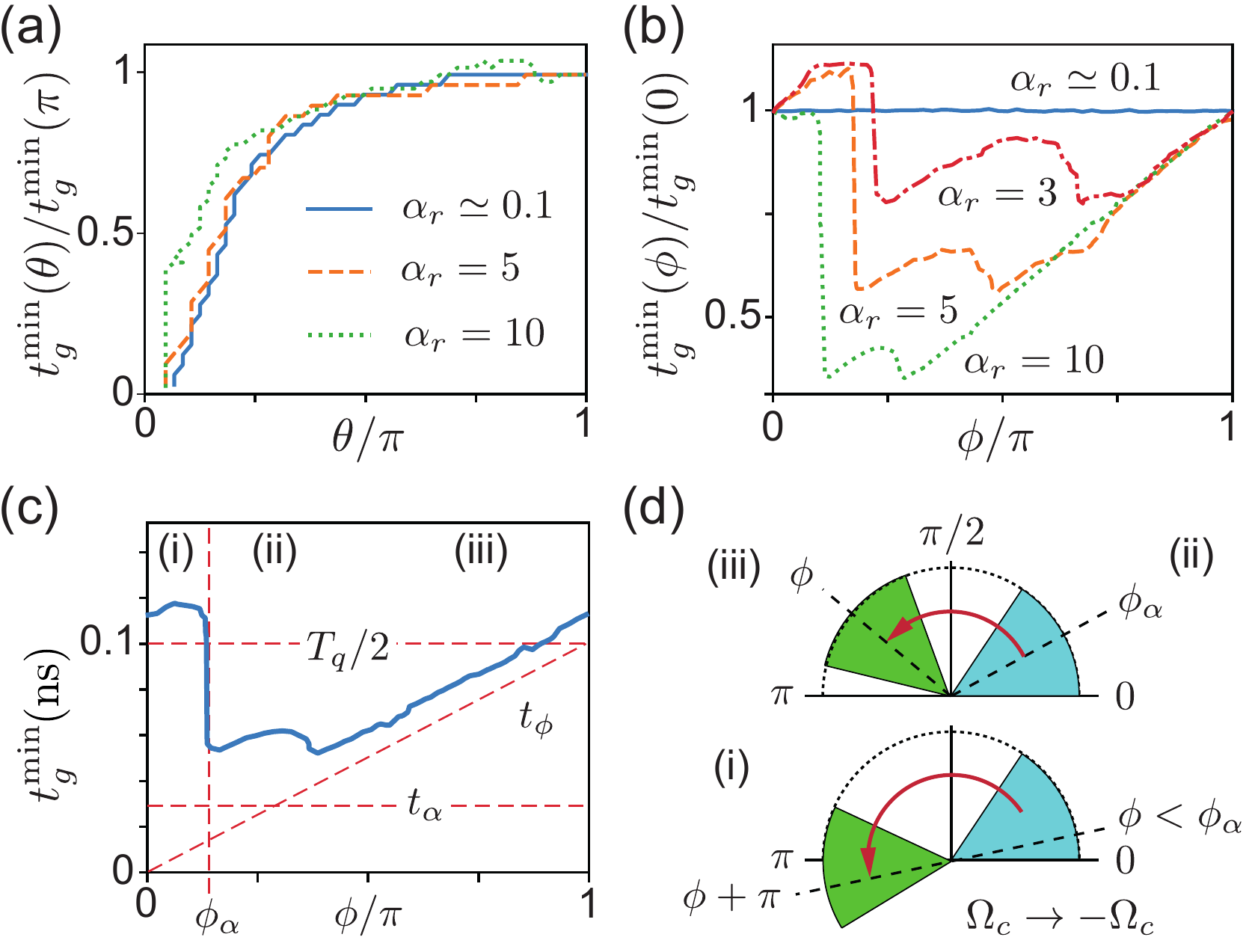}
    \caption{Minimal gate time $t_g^{\rm min}$ for different single-qubit rotations.  (a) Plot of the minimal gate time for realizing a $ R_Y(\theta) $ gate, normalized to $ t^{\rm min}_g(\theta = \pi) $. (b) Plot of the minimal gate time to perform a $ \theta =\pi/2 $ rotation about an axis $ n(\phi) = \cos(\phi) X + \sin(\phi) Y $, normalized to the minimal time of an $X$ rotation. The absolute gate times for the reference $R_X(\pi/2)$ rotations are $t_g=2.36$ ns ($\alpha_r=0.1$), $t_g=0.12$ ns ($\alpha_r=3$), $t_g=0.12$ ns ($\alpha_r=5$) and $t_g=0.11$ ns ($\alpha_r=10$). (c) Comparison of the minimal gate time for an $R_{n(\phi)}(\pi/2)$ rotation with the timescales $T_q=2\pi/\omega_{10}$, $t_\alpha=2\pi/|\alpha|$ and $t_\phi=\phi/\omega_{10}$, indicated by the dashed lines.  For this plot a large nonlinearity of $\alpha_r \simeq 7.3 $ has been assumed. (d) Interpretation of the control pulses obtained in the different regimes indicated in (c). See Sec.~\ref{subsec:RnRotations} for more details.  In all plots an error threshold of $\mathcal{E}_{\rm th}=10^{-4}$ is used for the definition of $t_g^{\rm min}$.}
    \label{fig:5_Angles}
\end{figure}

Let us now investigate more general single-qubit gates, considering first $R_Y(\theta)$ rotations with an arbitrary rotation angle $\theta$. In the limit of very large nonlinearities, all these rotations can in principle be implemented within the same time, by simply scaling the overall amplitude $\Omega_0$. Instead, when transitions to higher states become relevant, we expect that $\Omega_0$ should be kept approximately constant in order to reach the same fidelities as for the case of a $\pi/2$ rotation. This would correspond to a linear scaling $t_g(\theta)\approx \theta/\pi \times t_g(\pi)$. In Fig.~\ref{fig:5_Angles}(a) we plot the numerically minimized time $t^{\rm min}_g(\theta)$ that is required to implement $R_Y(\theta)$ rotations with an error below $\mathcal{E}_{\rm th}=10^{-4}$. We see that the actual dependence is between the two scenarios, but over the range of $\theta\in [\pi/4,\pi]$ the minimal gate time does not vary considerably. This means that the Hadamard gate $R_Y(\pi/2)$ discussed in detail above is already representative for most $R_Y$ rotations.

In next step we consider single-qubit gates of the form
\begin{equation}\label{eq:ShiftedYPulse}
R_{n(\phi)}(\theta)=\ex^{-\im\theta(\cos(\phi)\sigma^x+\sin(\phi)\sigma^y)/2},
\end{equation}
i.e., single-qubit rotations where the rotation axis, $ n(\phi) = \cos(\phi) X + \sin(\phi) Y$,  lies in the equatorial plane of the Bloch sphere. As discussed below Eq.~\eqref{eq:WeakDriving}, for weak driving fields all these gates can be implemented with the same pulse by simply setting the phase of the carrier to $\phi_d=-\phi$. In Fig.~\ref{fig:5_Angles}(b) we plot the numerically minimized gate time $t_g^{\rm min}(\phi)$ as a function of $\phi$ and find that it is indeed almost independent of $\phi$ for small nonlinearities $\alpha_r$. However, this is no longer true for large $\alpha_r$, where also rather abrupt jumps in the minimal gate time can be observed.

To explain these variations for ultrafast gates, we show in Fig.~\ref{fig:5_Angles}(c) a direct comparison between $t_g^{\rm min}(\phi)$ and other relevant timescales in this problem. We can identify three qualitatively different regions. For  the larger values of $\phi$ in region (iii) we find an almost linear relation, $t_g^{\rm min}(\phi)\sim \phi$, which can be understood as follows. Given a control pulse $\Omega_c^{\phi}(t)$ for implementing an $R_{n(\phi)}(\theta)$ rotation, the shifted pulse 
\begin{equation}\label{eq:ShiftedPulses}
\Omega_c^{\phi+\phi'}(t)=\Omega_c^{\phi}(t-t_{\phi'}),
\end{equation}
where $t_{\phi'}=(\phi'/2\pi)T_q$, realizes an equivalent rotation about the axis $n(\phi+\phi')$. This is a simple consequence of the fact that our qubit states are defined in a rotating frame and any shift of the pulse with respect to $t=0$ translates into a corresponding rotation in the $X-Y$ plane. Although the actual optimized control pulses in region (iii) of Fig.~\ref{fig:5_Angles}(c) are more complicated, they cannot outperform this simple waiting strategy, which is also  illustrated in the upper panel of Fig.~\ref{fig:5_Angles}(d). For very large $\alpha_r$ an accurate fit to the minimal gate times in region (iii) is given by $t_g^{\rm min}(\phi)\simeq t_{\phi}+\pi/|\alpha|$.  

For the $R_Y(\pi/2)$ rotation we have found above that the gate times are limited from below by $t_g > t_\alpha=  2\pi/|\alpha|$. This bound is set by transitions out of the qubit subspace and is thus expected to hold for any $R_{n(\phi)}$ gate. Although this bound is not fully reached for large $\alpha_r$, it still explains the  plateau for the minimal gate time observed in region (ii) of Fig.~\ref{fig:5_Angles}(c). Finally, for very small angles $\phi$, i.e., for a rotation axis close to the $X$ axis, we find a sharp jump of the minimal gate time to values $t_g^{\rm min}(\phi\approx 0)\gtrsim T_q/2$. The inability to implement faster $R_X$ rotations arises from the fact that during the minimal gate time $t_\alpha$ the average angle of the rotation axis is $\phi_\alpha=\pi t_\alpha/T_q$ [see Fig.~\ref{fig:5_Angles}(d)]. Therefore, for the same pulse duration, rotations about an axis with $\phi<\phi_\alpha$ become impossible. As indicated in the lower panel of Fig.~\ref{fig:5_Angles}(d), it is then the optimal strategy to find a suitable control pulse for an $R_{n(\phi+\pi)}$ rotation and use the inverted pulse $\Omega_c^{\phi}(t)=-\Omega_c^{\phi+\pi}(t)$ to implement the indented $R_{n(\phi)}$ gate. According to the waiting strategy discussed above, the total time for this pulse should be about $t_g\approx t_{\phi+\pi}$, consistent with the numerically optimized gate times. Note that flipping the sign of the control pulse can also be used to realize any other $R_{n(\phi)}$ gate with $\phi\in (\pi,2\pi]$ by simply inverting the corresponding control pulse $\Omega_c^{\phi-\pi}(t)$.

\subsection{Limit on the speed of single-qubit operations}

In summary we find that for the whole parameter range explored in our simulations, there exists a lower bound for implementing single-qubit rotations, 
\begin{equation}\label{eq:GateTimeLimit}
t_g^{\rm min} \gtrsim   \frac{2\pi}{|\alpha|} \times C(\theta,\mathcal{E}_{\rm th}),
\end{equation}
where for $\mathcal{E}_{\rm th}=10^{-4}-10^{-3}$ and for most rotation angles we can set $C(\theta,\mathcal{E}_{\rm th})\approx 1$. For small and moderate nonlinearities this bound is almost reached for all $R_{n(\phi)}$ rotations, while for very fast gates there appears another limitation from the qubit frequency,
\begin{equation}\label{eq:GateTimeLimit2}
t_g^{\rm min} (\phi) \gtrsim \frac{1}{\omega_{10}}  \times
\begin{cases}
\phi+\pi,\qquad \phi \in [0,\phi_\alpha),\\
\phi,\qquad  \,\,\,\,\,\,\,\,\,\,\,\phi\in [\phi_\alpha, \pi+\phi_\alpha),\\
\phi-\pi,\qquad \phi\in [\pi +\phi_\alpha, 2\pi).\\
\end{cases}
\end{equation}
Note that this second bound is less fundamental and could be avoided by implementing a two-axis control Hamiltonian. For example, by adding another driving  field that affects the charge variable, $H^\prime_{\rm drive}(t)\sim q$, one can choose the effective rotation axis at $t=0$ and therefore implement all $R_{n(\phi)}$ gates within the same time. However, adding one additional control line per qubit increases the circuit complexity and might not always be feasible. Therefore, here we restrict our analysis to a minimal circuit design. 

We remark that the bound in Eq.~\eqref{eq:GateTimeLimit} has already been previously discussed for weakly nonlinear three-level system under the validity of the rotating-wave approximation~\cite{Rebentrost2009,Safei2009}. It has successively been shown that under the same assumptions this bound can be surpassed using a two-axis control Hamiltonian, in which case essentially arbitrarily short gate times are possible~\cite{Motzoi2009}. While our optimization results confirm this conclusion for the three-level system, we do not find any considerable improvements of the minimal gate time when the full circuit Hamiltonian is taken into account. In our exact simulations the speed limit in Eq.~\eqref{eq:GateTimeLimit} is established as a lower bound for $t_g^{\rm min}$, both for single- and for two-axis control.

\subsection{Z rotations}\label{subsec:Zrotations}
Finally, let us briefly comment on the implementation of $Z$ rotations. In the single-qubit control Hamiltonian given in Eq.~\eqref{eq:h_drive} there is no term $\sim \sigma^z$ and we consider instead a quasi-adiabatic strategy for implementing a relative phase shift between the qubit states. To do so, the control field $\Omega(t)$ is slowly turned on and off again, such that the qubit follows adiabatically the rotated eigenstates of $H_q(t)$. In the limit of a two-level system, these instantaneous eigenstates are split by a time-dependent frequency
\begin{equation}
\omega_{10}^{\rm ad}(t) = \sqrt{\omega_{10}^2 + 4\varphi_{10}^2\Omega^2(t)}.
\end{equation} 
Therefore, this tunability can be used to implement an $R_Z(\theta)= e^{-\im\theta \sigma^z/2}$ gate with a rotation angle 
\begin{equation}
\theta = - \int_0^{t_g} dt' \, [\omega^{\rm ad}_{10}(t')- \omega_{10}],
\end{equation}
assuming that the system evolution remains fully adiabatic. 
\begin{figure}[b]
    \centering
    \includegraphics[width=\linewidth]{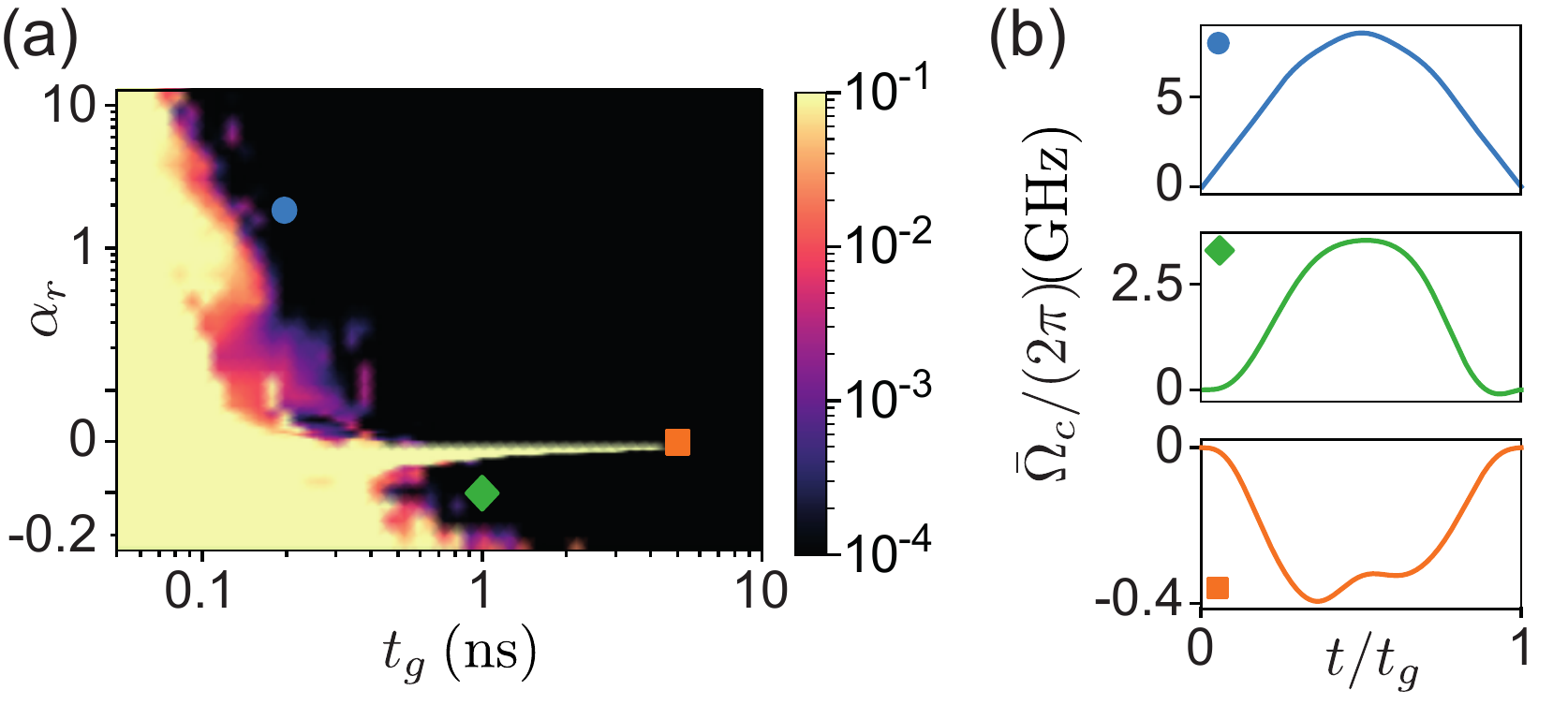}
    \caption{(a) Plot of the minimal gate error $\mathcal{E}$ for an $R_Z(\pi/2)$ rotation as a function of the qubit nonlinearity and the gate time. (b) Examples of optimized pulse shapes for $\bar \Omega_c(t)=\varphi_{10} \Omega_{c}(t)$ for the points marked in (a).}
    \label{fig:RZ}
\end{figure}

In Fig.~\ref{fig:RZ} we plot the results of a numerically optimized $R_Z(\pi/2)$ rotation, where we use the ansatz for $\Omega_c(t)$ given in Eq.~\eqref{eq:omega_c}, but without the carrier. This plot shows that compared to $R_{n(\phi)}$ rotations rather fast $R_Z$ gates can be implemented even with very small qubit nonlinearities. We attribute this different behavior to the fact that non-adiabatic state flips are suppressed by $\omega_{10}$ and not by $\alpha$. This also explains the rather sharp bound for high-fidelity gates around $t_g\approx \pi/\omega_{10}$, which depends only weakly on the value of $\alpha_r$. 

From this and other examples we conclude that down to gate times of $t^{\rm min}_g\gtrsim T_q$, fast and high-fidelity $R_Z$ rotations can be implemented without additional control terms in the circuit Hamiltonian.  Note, however, that the detailed findings for optimized $R_Z$ gates are much more sensitive to the precise value of the rotation angle and pulse optimization parameters and compared to other single-qubit gates no clearly interpretable trends for $t^{\rm min}_g$ are observed. For most applications this is not a relevant issue since in composite quantum circuits the $R_Z$ gates can either be eliminated completely or be replaced by an equivalent sequence of $R_{n(\phi)}$ gates. As we discuss in more detail in Appendix~\ref{app:ZRotations} this can be achieved with only a minor overhead on the total computation time such that the independent realization of $Z$ rotations can be advantageous, but is not strictly necessary.

\begin{figure*}[t]
    \begin{center}
	\includegraphics[width=0.9\textwidth]{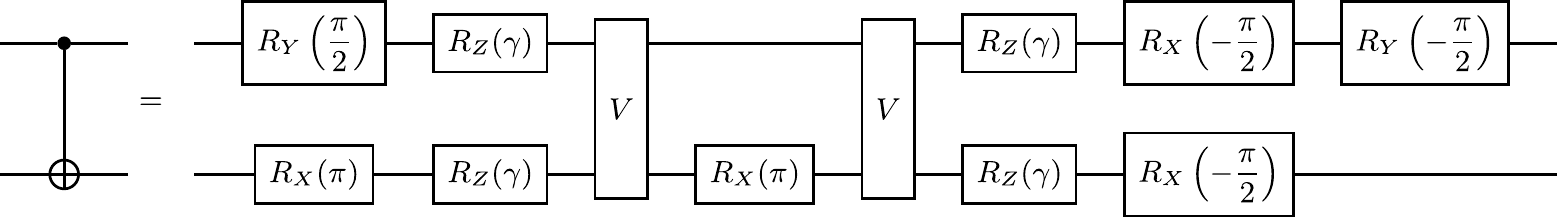}
	\caption{A possible decomposition of a CNOT gate into the native two-qubit gate $V$ and single-qubit rotations. The local phase $ \gamma $ is cancelled by $ R_Z(\gamma) $ gates. The $R_X(\pi) $ gate on the target qubit before and after the first $ V $ eliminates the conditional phase $ \beta $.}
        \label{fig:u_to_cnot}
    \end{center}
\end{figure*}

\section{Two-qubit gates}\label{sec:2q_gates}
In addition to single-qubit rotations, universal quantum computation requires at least one entangling two-qubit gate. To identify a suitable choice for this operation in the current setting we first consider the projection of $H_{qq}(t)$ in the qubit subspace, as given in Eq.~\eqref{eq:h2q}, and keep only energy conserving terms. Under these assumptions and for a gate time of $t_g = \pi/(4g_x)$ we obtain the time evolution operator 
\begin{align}\label{eq:V}
    V =
    \begin{pmatrix}
        \ex^{-\im\,(\beta - \gamma)} & 0 & 0 & 0\\
        0 & \ex^{\im\beta}/\sqrt{2} & -\im\ex^{\im\beta}/\sqrt{2} & 0\\
        0 & -\im\ex^{\im\beta}/\sqrt{2} & \ex^{\im\beta}/\sqrt{2} & 0\\
        0 & 0 & 0 & \ex^{-\im\,(\beta + \gamma)}
	\end{pmatrix},
\end{align}
where $ \gamma = \delta\omega t_g$ and $ \beta = g_z t_g $.  Up to single-qubit $R_Z$ rotations and an overall phase this gate is equivalent to the unitary
\begin{align}
    \tilde{V} =
    \begin{pmatrix}
        1 & 0 & 0 & 0\\
        0 & 1/\sqrt{2} & -\im/\sqrt{2} & 0\\
        0 & -\im/\sqrt{2} & 1/\sqrt{2} & 0\\
        0 & 0 & 0 & \ex^{-4\im\beta}
    \end{pmatrix},
\end{align}
which is the product of a $ \sqrt{\mathrm{iSWAP}} $ and a CPHASE~\cite{Krantz2019}. Note that $\tilde V$ is the same gate as implemented, for example, in the Sycamore quantum processor \cite{Arute2019,Foxen2020}, where it is also known as the `fSim gate' and used for quantum computation and quantum simulation applications. The $V$ gate can further be converted into a CNOT, as relevant for many circuits, by applying it twice and combining it with single-qubit rotations. Importantly, the corresponding sequence of gates shown in Fig.~\ref{fig:u_to_cnot} does not rely on any specific values for $ \gamma $ and $ \beta $. Therefore, this equivalence shows that $V$ is universal for \emph{arbitrary} values of $ \gamma $ and $ \beta $ and for the optimization of $V$, these parameters do not impose any constraints.

The qubit-qubit coupling $H_{qq}(t)$ can be controlled via the external flux $\Phi_s(t)= \pi\Phi_0 + \Phi_c(t)$ and thus our goal is to find a control pulse $\Phi_c(t)$ for which the actual unitary, 
\begin{equation}
U_2(t_g)= \mathcal{T} \ex^{-\im \int_0^{t_g} \id t \, \tilde H_{qq}(t)}, 
\end{equation}
approximates $V$ with the highest fidelity. For the numerical optimization we use a similar ansatz as for the single-qubit gates,
\begin{align}\label{eq:phi_c}
   	\Phi_c(t) = \Phi_{0}\sum_{n=1}^{n_{\rm max}} a_n\sin\left(\frac{ n\pi  t}{t_g}\right),
\end{align}
but without a carrier frequency. We also add the Josephson energy of the coupler, $ E_{J_s} $, as a static optimization parameter, which typically assumes values of about $ E_{J_s}/h \sim t^{-1}_g $. In contrast to the single-qubit gates, we now employ a slightly different cost function 
\begin{equation}\label{eq:cost_2q}
	\mathcal{C}_V = 1 - \dfrac{1}{d^2}| {\rm diag}[V^{\dagger}U_2(t_g)]|^2.
\end{equation}
Here $ d = 4 $ and the vector norm $|\cdot|$ is evaluated for the diagonal of the matrix product. This cost function is insensitive to the phases $ \gamma $ and $ \beta $, which is desirable since both can take arbitrary values. Note that this  cost function does not guarantee that $ U_2(t_g) $ will approach $ V $ as $ \mathcal{C}_V \to 0 $, since we lose information about the phases. However, it turns out that for the considered Hamiltonian in Eq.~\eqref{eq:h2q} this is not a problem and we find that the difference between the cost function $\mathcal{C}_V$ used for the optimization and the gate error $\mathcal{E}$ defined by Eq.~\eqref{eq:cost} is negligible. Importantly,  in all our plots we show the actual gate error $\mathcal{E}$.

\begin{figure}[t]
    \begin{center}
        \includegraphics[width=\columnwidth]{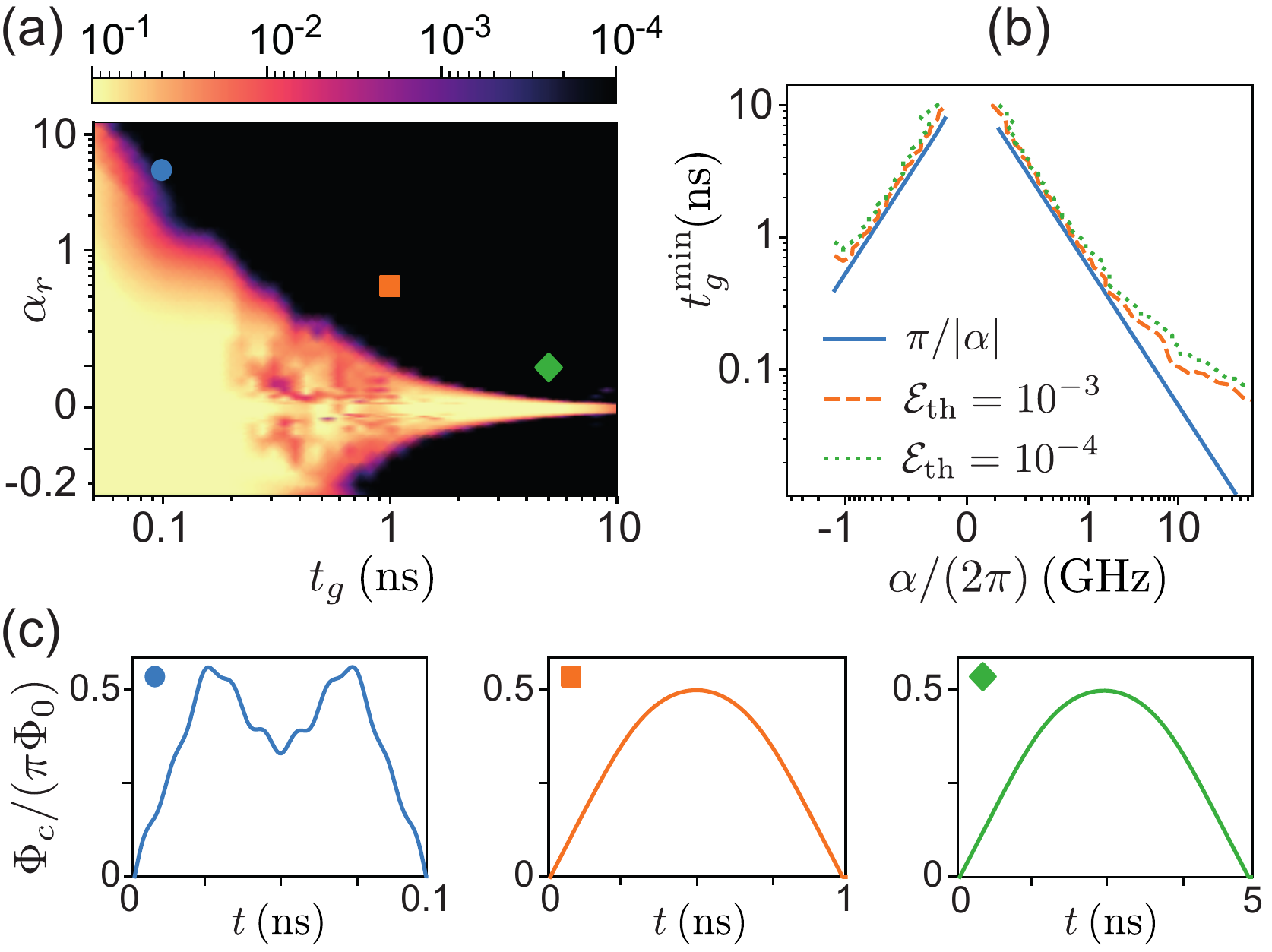}
        \caption{Optimization of the two-qubit gate $V$. (a) Plot of the minimal gate error $\mathcal{E}$ for varying gate times and nonlinearity parameters. (b) Minimal gate time $t_g^{\rm min}$ as a function of the absolute nonlinearity $\alpha=\alpha_r\omega_{10}$ and for two different error thresholds. (c) Examples of the optimized pulse shapes for the points marked in (a). Note that in (a) the scale of the $y$ axis and in (b) the scale of the $x$ axis is linear in the range $ [-0.1, 0.1] $ and logarithmic elsewhere.}
        \label{fig:7_V_optimization}
    \end{center}
\end{figure}

In Fig.~\ref{fig:7_V_optimization} we present the results for the numerically minimized errors for the two-qubit gate $V$, for varying gate times and nonlinearity parameters. Overall the results look very similar to what we have obtained for the single-qubit gates. Again, we find that high-fidelity two-qubits gates can be implemented on timescales $t_g\lesssim 100\,\mathrm{ps}$, if the qubit nonlinearity is sufficiently large. The corresponding optimal pulses for the control flux $\Phi_c(t)$ shown in Fig.~\ref{fig:7_V_optimization}(c) are rather smooth in all regimes.

An unexpected finding in the optimization of two-qubit gates is that for gate times in the range of $ \sim 0.3-1\,\mathrm{ns} $ there are individual points where very high gate fidelities can be reached for very low qubit nonlinearities. As discussed in more detail in Appendix~\ref{app:Outliers}, this observation can be explained by the fact that the coupling Hamiltonian $H_{qq}(t)$ is by itself nonlinear and can suppress excitations to higher states even if the qubit circuit is almost linear. However, since such exceptional cases are rather rare and are no longer found for the bandwidth-limited pulses discussed below, we do not go into possible re-definitions of the nonlinearity parameter to account for such conditions here. Therefore, leaving these fine-tuned outliers aside, we plot in Fig.~\ref{fig:7_V_optimization}(b) the minimal gate time $t_{g}^{\rm min}(\alpha)$ for the two-qubit gate as a function of the absolute single-qubit nonlinearity. From this plot we extract the lower bound
\begin{equation}
t_g^{\rm min}(\alpha)>  \frac{\pi}{|\alpha|}.
\end{equation} 
We see that the optimized pulses approach this bound for a large range of nonlinearities. Again, we find a deviation from this scaling for gate times of $t_g \lesssim 0.1 \, {\rm ns}$, which shows that in this regime the assumed control Hamiltonian $\sim XX$ is no longer optimal to implement a gate derived under the rotating-wave approximation. This problem can be overcome by implementing a more flexible coupling Hamiltonian or by optimizing not for a predefined, but for an $\alpha_r$-specific entangling operations, similar to what has been discussed in Refs.~\cite{Watts2015,Goerz2015b}.


\section{Bandwidth limitations}\label{sec:bw}
In all the examples so far we considered control pulses of essentially arbitrary shape. Although arbitrary waveform generators (AWGs) with sampling rates of $ 50\,\mathrm{GHz} $ and more are commercially available, for ultrafast gates the corresponding discretization steps of $\sim 20\,\mathrm{ps}$ are still comparable to the total gate time. In addition, the rise and fall time between the voltage steps are finite and the bandwidth of the actual output signal of an AWG is in general smaller than the sampling rate. The propagation of the signal through attenuators and cables can lead to further pulse dispersion and filtering effects. Therefore, in the picosecond domain, control pulses of arbitrary shape are no longer available and bandwidth limitations become one of the major technical limitations for implementing fast and high-fidelity quantum gates.

To account for finite-bandwidth effects without going into the details of the control electronics, we here simply assume that the driving field $\Omega(t)$ for the qubit is related to the control pulse $\Omega_c(t)$ via the linear transformation
\begin{equation}
\Omega(t) = \int_{-\infty}^t \id t' \, F(t-t') \Omega_{c}(t'),
\end{equation}
and a corresponding relation for $\Phi_s(t)$ and $\Phi_c(t)$. Here $F(t)=\int_{-\infty}^{\infty} \id\omega \, \ex^{\im\omega t} F(\omega)$ is a filter function, which is setup-specific. For concreteness, we consider in all our calculations a Butterworth filter \cite{Butterworth1930} with a frequency response
\begin{equation}
    F(\omega) = \dfrac{\im\omega}{\prod_{k = 1}^{m} \left[ \omega - \im\Delta_F\ex^{-\im\pi(2k + m - 1)/(2m)} \right]},
\end{equation} 
where $m$ is the order and $\Delta_F$ is the bandwidth of the filter. In Fig.~\ref{fig:8_Filter}(a) we illustrate the effect of this filter for $m=4$ and $\Delta_F/(2\pi) = 15\,\mathrm{GHz}$. The plot shows the original control signal $\Omega_c(t)$ derived above for a gate time of $t_g=0.2\,\mathrm{ns}$ and the resulting driving field for the qubit, $\Omega(t)$. We see that the filter not only distorts the pulse, but it also induces a significant delay and a non-vanishing driving signal for times $t>t_g$.

\subsection{Optimization of filtered control pulses}
\begin{figure}[t]
    \begin{center}
        \includegraphics[width=\columnwidth]{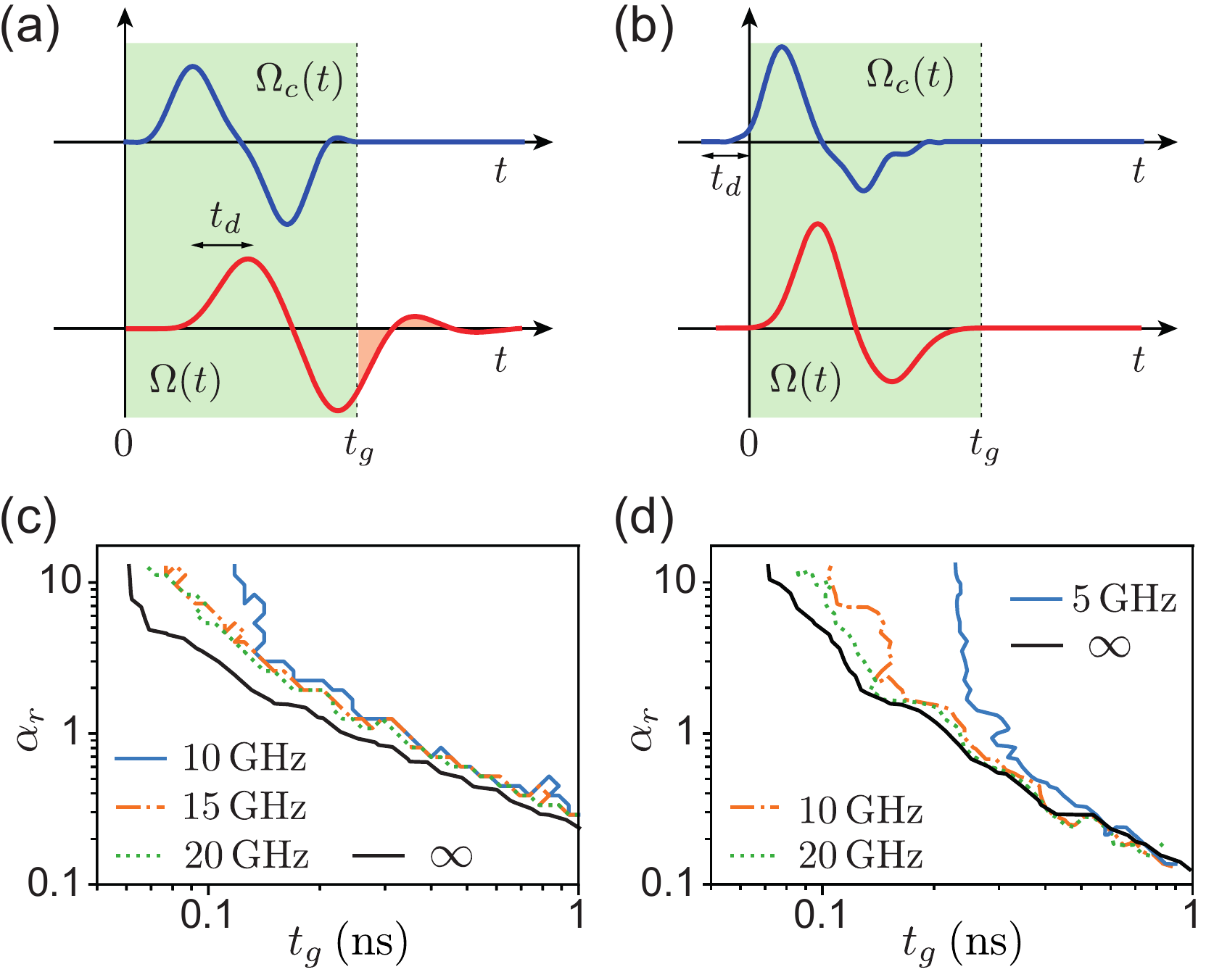}
	\caption{Optimization of the single- and two-qubit gates for control pulses with limited bandwidth. (a) When the original control signal, $\Omega_c(t)$,  is passed through a lowpass filter, the pulse driving the qubit, $\Omega(t)$, gets distorted, delayed and has a significant support outside the targeted gate interval, $[0,t_g]$. (b) To compensate for these effects, the control signal is shifted and re-optimized, such that the filtered pulse minimizes the total cost function $\mathcal{C}$ defined in Eq.~\eqref{eq:1q_penalty} for single-qubit gates and Eq.~\eqref{eq:2q_penalty} for two-qubit gates. (c) and (d) Minimized total gate error $ \mathcal{E}_F$ for implementing (c) an $R_Y(\pi/2)$ rotation and (d) a $V$ gate with bandwidth-limited driving fields. The contour lines indicate the error threshold $ \mathcal{E}_{\rm th} = 10^{-4} $ for a Butterworth filter of order $m=4$ and different bandwidths $ \Delta_F $. The black solid lines represent the same error threshold for the unfiltered control pulses taken from Fig.~\ref{fig:3_H_Optimization} and Fig.~\ref{fig:7_V_optimization}. Note that for the filtered pulses in (d) we have doubled the amount of iterations in the numerical optimization.} 
        \label{fig:8_Filter}
    \end{center}
\end{figure}

For the optimization of bandwidth-limited control pulses, let us first of all emphasize that our goal is to implement a target single- or two-qubit operation within the pre-specified time interval $[0,t_g]$. The example in Fig.~\ref{fig:8_Filter}(a) shows that for filtered pulses a non-negligible part of the qubit evolution can take place outside the intended gate interval. Therefore, in this case not only the optimization procedure, but also the definition of gate errors must be adapted. 

To compensate for the filter-induced delay we can simply start the control pulse a little earlier, i.e., $\Omega_c(t)\rightarrow \Omega_c(t+t_d)$, where $t_d$ is adapted during the optimization process. As before, we then calculate the fidelity of the resulting unitary evolution during the time interval $[0,t_g]$ and denote the corresponding gate error by $\mathcal{E}_{[0,t_g]}$. However, this error does not take into account that also for times $-t_d<t<0$ and $t>t_g$, the driving field does not vanish completely and can induce additional rotations of the qubit state. To estimate these errors we calculate as well the circuit evolution over a longer time interval, 
\begin{equation}
U_g^\prime= \mathcal{T} \ex^{-\im \int_{t_i}^{t_f} \id t \, \tilde H_{\rm drive}(t)},
\end{equation}
and evaluate the corresponding error $\mathcal{E}_{[t_i,t_f]}$. In our simulations we choose $t_i=-t_d$ and $t_f=10t_g$, but we have verified that the results do not change considerably when the length of this interval is varied. Since we can improve $\mathcal{E}_{[0,t_g]}$ at the expense of $\mathcal{E}_{[t_i,t_f]}$ and vice versa, we define the gate error for filtered pulses as 
\begin{equation}\label{eq:EF}
\mathcal{E}_F={\rm max}\{ \mathcal{E}_{[0,t_g]},\mathcal{E}_{[t_i,t_f]}\}.
\end{equation}
In Fig.~\ref{fig:8_Filter}(b) we show an example of a control pulse $\Omega_c(t)$, which minimizes this generalized gate error, $\mathcal{E}_F$ (see Appendix \ref{app:optimization} for further details). While in this case the control signal can be a bit more complicated, the actual filtered signal that drives the qubit is rather smooth and most of its support is contained in the targeted time interval $[0,t_g]$. 

\subsection{Bandwidth limitations for single- and two-qubit gates}

In Fig.~\ref{fig:8_Filter}(c) and Fig.~\ref{fig:8_Filter}(d) we repeated the optimization of the single- and two-qubit gates discussed in Fig.~\ref{fig:3_H_Optimization} and Fig.~\ref{fig:7_V_optimization} for different values of the filter bandwidth $\Delta_F$. For clarity, the plots show only the contour line at $\mathcal{E}_F=10^{-4}$. For small and moderate nonlinearities and the considered values of $\Delta_F$ we find no significant influence of the filter beyond small variations from the optimization procedure.  At intermediate values of $t_g$ the presence of the filter starts to degrade the achievable gate fidelities, but this can still be compensated by working with qubits with a slightly higher nonlinearity. Finally, depending on the value of $\Delta_F$ we find a rather sharp boundary, below which high-fidelity gate operations are no longer possible, independent of the degree of nonlinearity. This boundary is roughly consistent with $t_g^{\rm min}(\alpha_r\gg1)\approx 2\pi/\Delta_F$, as expected from the general time-frequency uncertainty relation.  This scaling indicates that the leakage of the driving signal outside the time interval $[0,t_g]$ is more detrimental than the elimination of some of the faster wiggles in the pulse.

\section{Ultrafast quantum circuits}
\label{sec:arb_circ}

The ability to realize any $R_{n(\phi)}(\theta)$ rotation together with the two-qubit unitary $V$ is in principle enough to construct arbitrary quantum circuits~\cite{NielsenChuang}.  For this reason, most optimization studies focus on the implementation of either individual or similar universal sets~\cite{Goerz2017,Abdelhafez2020} of single- and two-qubit gates. However, if one is interested in absolute processing times, there are several additional aspects and physical constraints that need to be taken into account when combining those individual gates into larger circuits. In this section we address, first of all, the clocking requirements for ultrafast quantum gates, which are imposed in the picosecond regime by a finite qubit rotation time. In a second step, we then describe the implementation of larger quantum circuits using an explicit example.

\subsection{Clocking of composite circuits}
In Sec.~\ref{sec:1q_gates} and Sec.~\ref{sec:2q_gates} we have optimized all our gate operations for the time interval $[0,t_g]$. Since our qubit states are defined in a rotating frame [see Eq.~\eqref{eq:Qubits}] also the meaning of $R_X$ and $R_Y$ gates is defined only with respect to this origin in time. Consequently, by applying the same control pulses during a different time interval $[t,t+t_g]$, the resulting gate operation will in general not be the same, except when $t$ is a multiple of the qubit precession  time $T_q=2\pi/\omega_{10}$. For single-qubit gates we can further reverse the sign of the control field to convert, for example, a rotation around $-X$ into a rotation around $X$ and thereby obtain identical gates already after a period of $T_q/2$. The same turns out to be true for the two-qubit gate $V$ given in Eq.~\eqref{eq:V}, since the period of the product $\sigma_1^x\sigma_2^x$ is half of the single-qubit oscillation time.  Therefore, we identify a minimal cycle time, $t_{\rm cyc}=T_q/2$, according to which all the control pulses must be clocked. 

\begin{figure}[t]
    \centering
    \includegraphics[width=\linewidth]{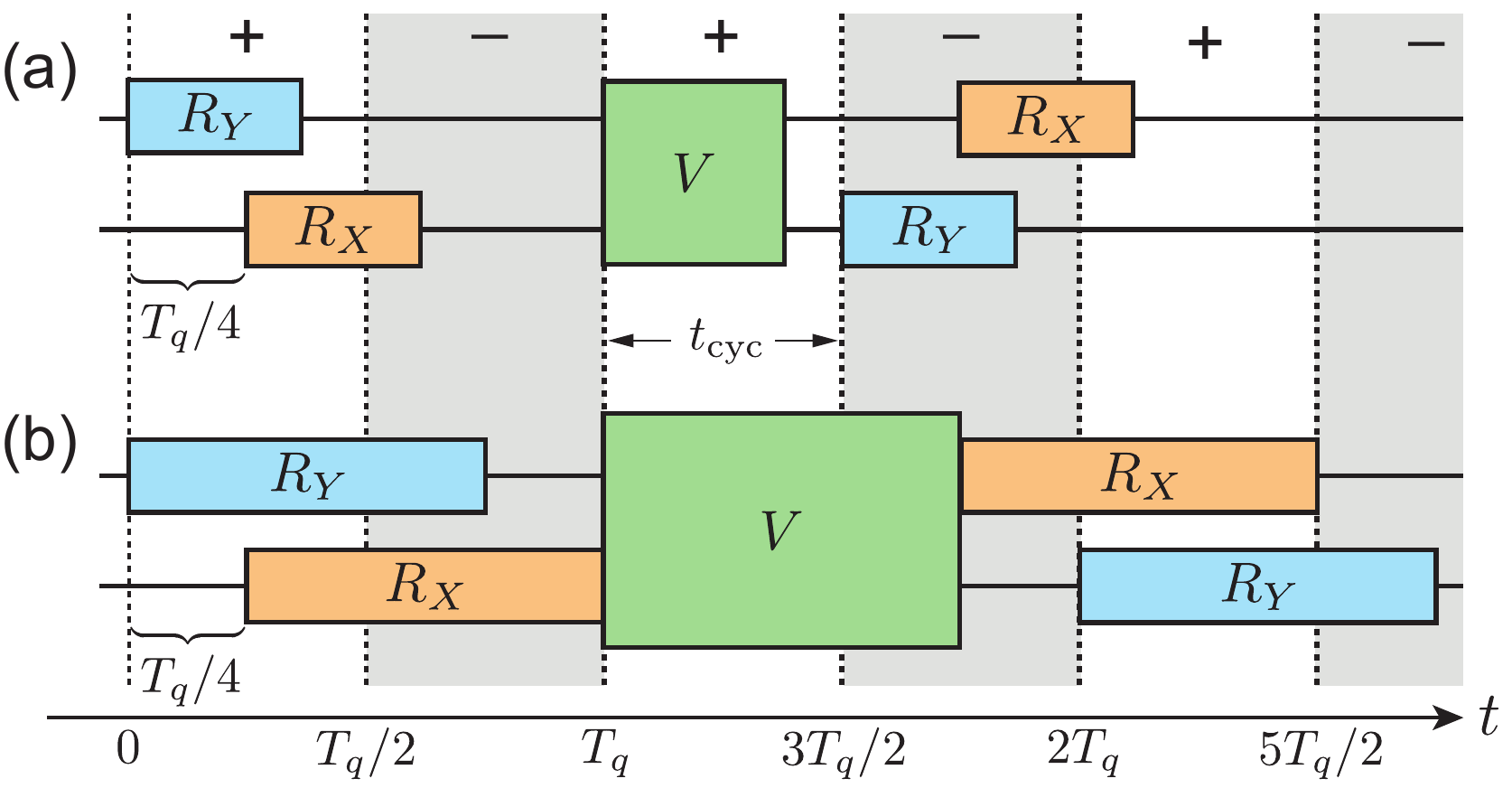}
    \caption{Constructing quantum circuits with ultrafast gates. The optimized gates are arranged according to a clocking cycle with period $t_{\rm cyc}=T_q/2$. The $R_X$ gates are shifted with respect to $R_Y$ gates by a waiting time $T_q/4$, as explained in Sec.~\ref{subsec:RnRotations}.  After each cycle the same single-qubit gates can be implemented by simply flipping the sign of the control pulse. For these examples it is assumed that the $R_Y$ and $V$ gates can be implemented in the same time $t_g$, where $t_g/t_{\rm cyc}=0.75 $ in (a) and $t_g/t_{\rm cyc}=1.5$ in (b). }
    \label{fig:9ClockCycle}
\end{figure}

In Fig.~\ref{fig:9ClockCycle} we illustrate how a finite clocking interval, which is $t_{\rm cyc}=0.1\,\mathrm{ns}$ in all our examples, influences the consecutive execution of ultrafast gates. For the two examples we assume the same gate time $t_g$ for implementing an $R_Y(\pi/2)$ rotation and the $V$ gate. As discussed in Sec. \ref{subsec:RnRotations}, an $R_X(\pi/2)$ gate can then be implemented with the same pulse shifted by $t_{\pi/2}=T_q/4$. We see that although in the first example $t_g=0.75 t_{\rm cyc}$ is two times shorter than in the second example, where $t_g=1.5 t_{\rm cyc}$, the run time of the whole circuit is not too much different in the two cases. This illustrates that for ultrafast gate operations the bare qubit frequency $\omega_{10}$ can set an important limitation for the overall computation speed. Note that $R_Z$ rotations are invariant under a shift of the phase of the qubit states and can thus be implemented at any time.

\begin{figure*}[t]
    \begin{center}
	    \includegraphics[width=\textwidth]{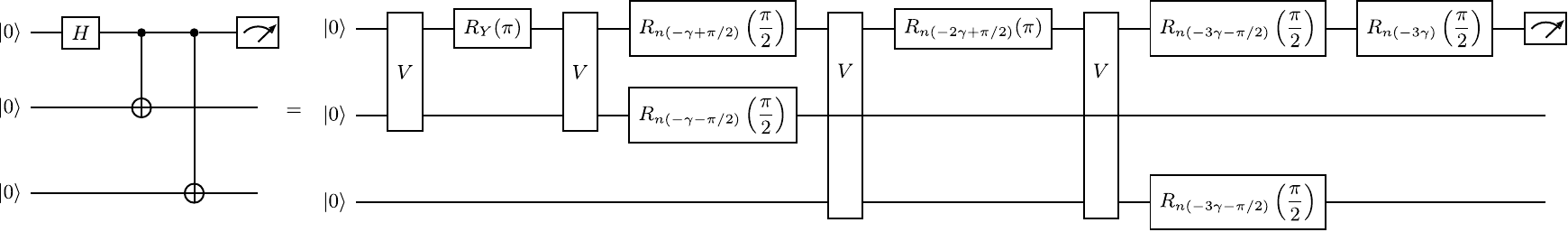}
	    \caption{The quantum part of a compressed Shor algorithm for factoring $ N = 15 $~\cite{Monz2016}. The left side shows the original circuit composed out of one Hadamard gate and two CNOT gates. The right side shows the equivalent circuit implemented with $V$ and $R_{n(\phi)}$ gates.}
    \label{fig:shor_demo}
    \end{center}
\end{figure*}

\subsection{Shor's algorithm in a nanosecond}\label{subsec:shor}

In a final step we now combine all the results and considerations from the previous sections and discuss an ultrafast implementation of a small composite quantum circuit. As an illustrative example, we consider here the three-qubit circuit shown in Fig.~\ref{fig:shor_demo}, which consists of one Hadamard and two CNOT gates. This particular circuit represents the most essential part in the implementation of a compressed version of Shor's algorithm for factoring the number 15.  For a more detailed discussion of this circuit and its relation to the original Shor algorithm~\cite{Shor1994,Shor1997} we refer the reader to Ref.~\cite{Monz2016}, where the implementation of the same algorithm has been demonstrated with trapped ions. For the current purpose it is enough that this circuit executes a minimal useful quantum computation, but at the same time it still permits an exact numerical simulation of the full superconducting circuit that is used to encode the three qubits.

From a naive decomposition of the two CNOT gates into $V$ and single-qubit operations, as shown in Fig.~\ref{fig:u_to_cnot}, we would obtain $25$ elementary gates and a total execution time of about $T_{\rm circ}\approx 20 t_g$. However, by implementing further simplification we end up with the equivalent circuit shown in Fig.~\ref{fig:shor_demo}, which is reduced to four applications of $V$ and seven single-qubit rotations. Note that this circuit is equivalent to the original one up to $ Z $ rotations of the initial and final states.

\begin{figure}[h]
	\centering
	\includegraphics[width=\columnwidth]{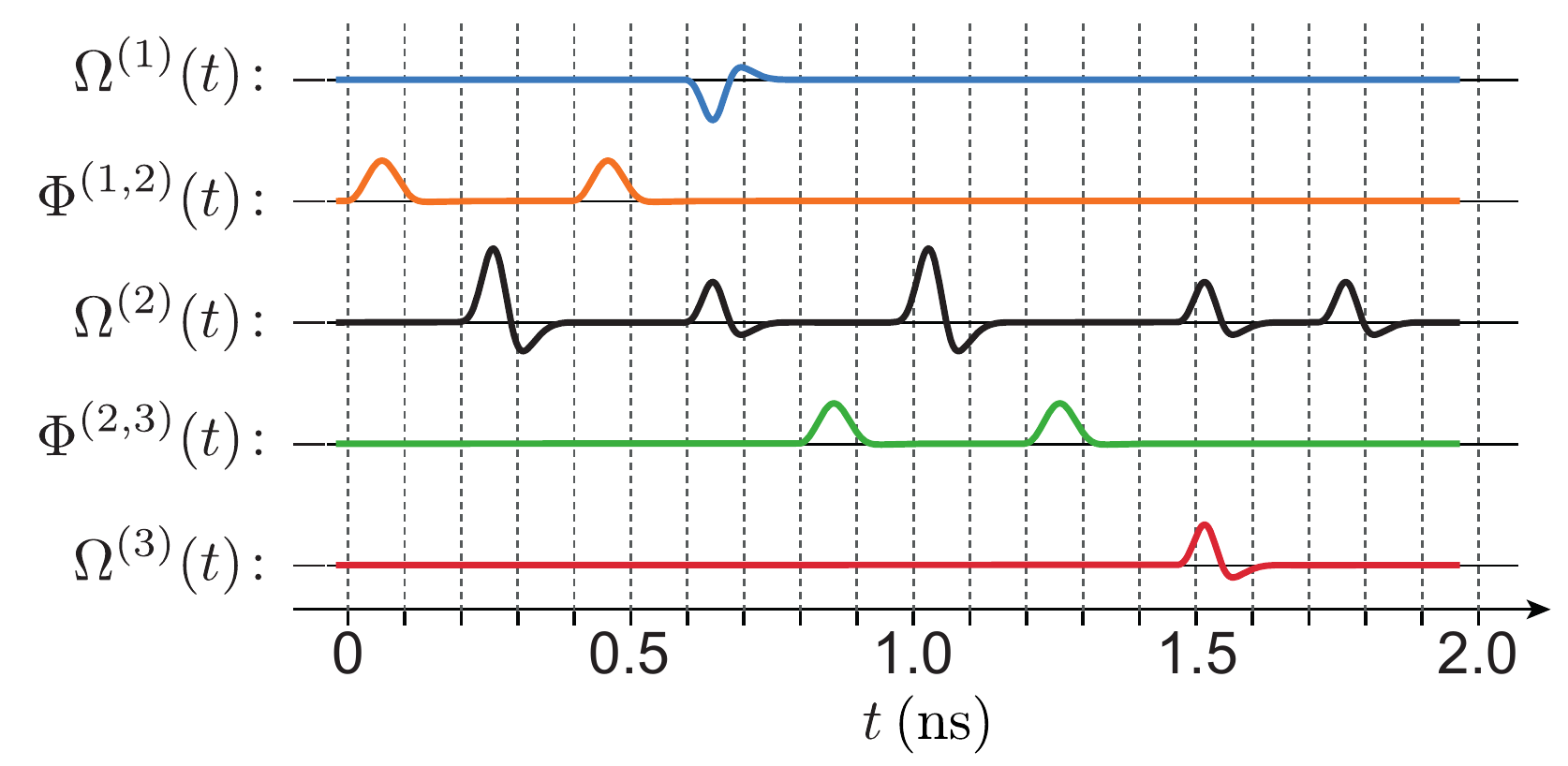}
    \caption{Example of a pulse sequence for implementing the quantum circuit shown in Fig.~\ref{fig:shor_demo}. The five lines indicate the time dependence of the individual control fields $\Omega^{(i)}(t)$ for each qubit and the fluxes $\Phi^{i,i+1}(t)$ (omitting a constant offset $\pi\Phi_0$) for controlling the two-qubit interactions. Note that here qubit 2 is used as the control qubit such that the circuit can be implemented using only nearest-neighbor interactions. The parameters for these pulses are taken from example (ii) in Table~\ref{tab:ShorParameters}.}
    \label{fig:11_pulse_seq}
\end{figure}

To simulate the implementation of this circuit with superconducting qubits, we consider the full Hamiltonian 
\begin{equation}
H(t)= \sum_{i=1}^3 H_q^{(i)}(t) + H_{qq}^{(1,2)}(t)+ H_{qq}^{(2,3)}(t),
\end{equation} 
as described in Sec.~\ref{sec:model}. For a given nonlinearity parameter $\alpha_r$ and filter bandwidth $\Delta_F$, we then find numerically optimized control pulses for implementing individual $R_{Y}(\pi/2)$ and $R_Y(\pi)$ rotations and the $V$ gate. For all three gates we assume the same time $t_g$, which is chosen such that $\mathcal{E}_F \lesssim 10^{-4}$ for each individual gate. All the remaining single-qubit gates are then implemented using the same, but shifted control pulses, as described in Eq.~\eqref{eq:ShiftedPulses}. In cases where there is no overlap with a preceding gate, the time shifts $t_\phi$ can also be chosen negative to avoid unnecessary waiting periods. All these individual pulses are then matched with the clocking cycle and combined into a full control sequence, as described in Sec.~\ref{sec:arb_circ} A. An example for the resulting control signals for the three qubits and the two couplers is shown in Fig.~\ref{fig:11_pulse_seq}. Note that this procedure of combining control pulses is not fully optimal, but it makes use of only on a small number of optimized gates and does not rely on any specific details of the circuit.

Following this procedure, we simulate the implementation of the full quantum circuit for different combinations of $\alpha_r$ and $\Delta_F$. The detailed set of gate parameters is summarized in Table~\ref{tab:ShorParameters}. For the full simulation we initialize the qubits in the state $|\Psi(0)\rangle =|0\rangle_1|0\rangle_2|0\rangle_3$ and calculate the final state 
\begin{equation}
|\Psi(T_{\rm circ})\rangle = \mathcal{T} \ex^{-\im \int_0^{T_{\rm circ}} \id t \, \tilde H(t)} |\Psi(0)\rangle,
\end{equation}
including the $10$ lowest basis states for each subcircuit.  We define the error of the full computation in terms of the state overlap,
\begin{equation}
\mathcal{E}_{\rm circ}= 1- |\langle \Psi_{\rm target}| \Psi(T_{\rm circ})\rangle|^2,
\end{equation}
 where $|\Psi_{\rm target}\rangle$ is the targeted state in the qubit subspace.  
 
 \begin{table}[t]
	\centering
    \bgroup
    \def\arraystretch{1.6}
	\begin{tabular}{l|c|c|c}
	 & (i) & (ii) & (iii) \\
	 \hline
         $ \alpha_r $ & -0.05 & 5 & 10 \\
        \hline
         $ \Delta_F/(2\pi) $ & $ 0.75\,\mathrm{GHz}^*  $ & $ 10\,\mathrm{GHz} $ & $ 15\,\mathrm{GHz}  $ \\
        \hline
             $ t_g $ & $ 6\,\mathrm{ns} $ & $ 170\,\mathrm{ps} $ & $  100\,\mathrm{ps}$ \\
        \hline
        $ \gamma $ & $-0.5039 $ & $ 0.4594 $ & $  0.5863$ \\
        \hline
        $ \mathcal{E}_F^{\pi/2} $ & $\,\,\, 4.4\times10^{-5} \,\,\, $ & $\,\,\, 6.0 \times 10^{-6} \,\,\, $ & $ \,\,\, 3.9\times 10^{-4} \,\,\, $ \\
        \hline
        $ \mathcal{E}_F^{\pi} $ & $ 8.7\times10^{-5}$ & $ 9.9\times 10^{-5} $ & $ 1.6\times 10^{-4} $ \\
        \hline
	$ \mathcal{E}_F^{V} $ & $3.8\times 10^{-7} $ & $ 5.0\times 10^{-6} $ & $ 3.9\times 10^{-5} $ \\
        \hline
        $ T_{\rm circ} $ & $ 54.3\,\mathrm{ns}  $ & $ 1.9\,\mathrm{ns} $ & $1.2\,\mathrm{ns}   $ \\
        \hline
        $ \mathcal{F}_{\rm circ} $ & $ 0.9996 $ & $ 0.9998 $ & $ 0.9984 $
	\end{tabular}
    \egroup
    \caption{Summary of the parameters used for the simulation of the quantum circuit shown in Fig.~\ref{fig:shor_demo}. The different $\mathcal{E}_F$ denote the gate error define in Eq.~\eqref{eq:EF} for the $R_Y(\pi/2)$, the $R_Y(\pi)$ and the $V$ gate, which are all implemented in the same time $t_g$. Note that in example (i) the filter is only applied to the pulse envelop and not the carrier.} 
	\label{tab:ShorParameters}
\end{table}

In the first example listed in Table~\ref{tab:ShorParameters}  we consider a conventional transmon qubit with a nonlinearity parameter of $\alpha_r=-0.05$. In this case we can choose a gate time of $t=6\,\mathrm{ns}$, close to the bound in Eq.~\eqref{eq:GateTimeLimit}, and obtain a total execution time of $T_{\rm circ}\simeq  54\,\mathrm{ns}$. Note that this time for implementing the whole circuit is comparable or even faster than most of the individual gates that are used in experiments today \cite{Kelly2015,Kjaergaard2020,Jurcevic2020}. For the second example we choose a highly nonlinear qubit with $\alpha_r=5$ and $\Delta_F/(2\pi)= 10\,\mathrm{GHz}$. For these parameters and a gate time of $t_g=170\,\mathrm{ps}$ we obtain $T_{\rm circ}\simeq  1.9\,\mathrm{ns}$ with a total error $\mathcal{E}_{\rm circ}\simeq 2\times 10^{-4}$. Finally, in the third example we consider gates of only $t_g=100\,\mathrm{ps}$, assuming $\alpha_r=10$ and $\Delta_F/(2\pi)= 15\,\mathrm{GHz}$. In this case the total algorithm can be implemented in just above $ 1\,\mathrm{ns} $, still retaining a total error of about $\mathcal{E}_{\rm circ}\approx 2\times 10^{-3}$.

\section{Discussion and Conclusions}\label{sec:Conclusions and outlook}
In summary, we have presented a systematic study about the implementation of ultrafast quantum gates with superconducting circuits. In particular, by assuming a very generic qubit design, we have investigated the dependence of the minimal gate time on the qubit nonlinearity and the bandwidth of the control pulse over a large parameter range. Our numerical results show that there exists a lower bound of $t_g^{\rm min}\approx 2\pi/|\alpha|$ for single-qubit gates and $t_g^{\rm min}\approx \pi/|\alpha|$ for the considered two-qubit gate. This contradicts previous conclusions drawn from the optimization of three-level systems, where in principle arbitrarily fast gates can be implemented~\cite{Motzoi2009}. At the same time, over the whole range of parameters explored in this work, this bound does not depend on the precise level structure of the qubit circuit and is thus expected to apply also for all other qubit designs in use today.  Although based on purely numerical observations, the ansatz for the control pulses assumed in this work is completely generic with an exhaustive number of variational parameters. Since also no variation of the optimization strategy or the initial conditions led to a different result, we conclude that the observed bound represents indeed a fundamental limit for the gate time.
For very fast gates, $t_g\sim 100\,\mathrm{ps}$, we have found that additional restrictions arise from the finite qubit oscillation time, $T_q$, which, however, can in principle be overcome by changing, for example, to a two-axis control scheme.        

In the second part of this paper we have addressed in more detail the implementation of larger quantum circuits composed out of many ultrafast gates. Here again we have found that the finite qubit rotation time must be taken into account and introduces a natural cycle time $t_{\rm cyc}=T_q/2$ according to which gates must be clocked. We have illustrated this circuit composition by performing a full multi-level simulation of a basic three-qubit circuit consisting of eleven elementary single- and two-qubit gates. For realistic qubit nonlinearities and control bandwidths, the simulated execution times for the whole circuit are about $T_{\rm circ}\sim1-2\,\mathrm{ns}$. This is about hundred times faster than what is achievable in most superconducting quantum computing experiments today and demonstrate that significant improvements in this direction are still possible.  

In our analysis we have restricted ourselves to gate times down to about $t_g\sim 50\,\mathrm{ps}$, which require absolute nonlinearities of $\alpha/(2\pi)\gtrsim 25\,\mathrm{GHz}$ and even a bit larger control bandwidths. While such parameters are highly non-standard for current superconducting qubit experiments, they are still within physical and technological bounds. In particular, in this regime the involved frequencies remain below the value of $2\Delta_{\rm SC}\approx 2\pi\times 80\,\mathrm{GHz}$, which is twice the superconducting gap of aluminium, such that resonant excitations of quasi-particles are still negligible. 

Quasi-particles, which affect the coherence of the qubit, can also be generated through nonlinear processes at large driving strengths. Note, however, that while the maximal amplitude of the control pulse increases as $\Omega_0\sim 1/t_{\rm g}$, this scaling does not necessarily translate into an equivalent increase in the signal power. For example, all the two-qubit gates in Fig.~\ref{fig:7_V_optimization} are implement with approximately the same maximal value of the control flux. Also the mutual inductance for the single-qubit control can be effectively realized through a weakly driven Josephson junction and we remark that driving amplitudes of about $\Omega_0/(2\pi)\approx 5$ GHz have been experimentally demonstrated without a drastic impact on the qubit coherence~\cite{Deng2015}. While a detailed evaluation of the effect of quasi-particle production will be setup-specific and is beyond the scope of the current analysis, it is not expected to represent a major technical obstacle down to gate times of about $t_g\sim 100\,\mathrm{ps}$.

Finally, let us remark that while for the implementation of even faster gates the superconducting gap can no longer be ignored,  the gap can be substantially higher in other materials, such as niobium. This means that coherent operations of superconducting qubits on timescales as short as $t_g\approx 1-10\,\mathrm{ps}$ are at least physically still conceivable.

\acknowledgments
This work was supported by the National Natural Science Foundation of China (Grants No. 61675007 and No. 11975026), the Beijing Natural Science Foundation (Grant No. Z190005), the Austrian Science Fund (FWF) through Grant No. P31701 (ULMAC) and the European Union's Horizon 2020 research and innovation
programme under grant agreement No. 899354 (SuperQuLAN). D.Z acknowledges the financial support provided by China Scholarship Council (Grant No. 201906010204). 
\appendix

\section{Pulse optimization}\label{app:optimization}
To identify the optimal pulses for implementing single- and two-qubit gates, we parametrize the control pulses in terms of a finite set of variables and identify an appropriate cost function $\mathcal{C}$, which we want to minimize. For such generic optimization problems there are several standard numerical methods available and different versions of such algorithms have already been implemented in the past for optimizing quantum gates~\cite{Spoerl2007,Montangero2007,Motzoi2009,Rebentrost2009,Safei2009,Lucero2010,Gambetta2010,Egger2014,Schutjens2013,Motzoi2013,Huang2014,Theis2016,Liebermann2016,Kirchhoff2018,Theis2018,Machnes2018,GarciaRipoll2020,Xu2020,Watts2015,Goerz2015b,Glaser2015,Leung2017,Goerz2017,Abdelhafez2020,Shillito2020,Abdelhafez2019,Tian2020}. However, to obtain a sufficiently fast convergence, smooth pulse shapes, etc., usually a problem-specific tuning of these algorithms is required. In this appendix we summarize the detailed optimization procedure that has been used to produce all the results presented in this paper. 

\subsection{Pulse parametrization}
For the control pulses $\Omega_c(t)$ and $\Phi_c(t)$ we use the expansion in terms of sine waves as given in Eq.~\eqref{eq:omega_c} and Eq.~\eqref{eq:phi_c} and the pulse amplitude is set to zero for times outside the interval $[0,t_g]$. For all results we take the same number of frequency components $n_{\rm max}=20$ to represent the pulse envelope, which for the single-qubit gates is multiplied by a carrier wave of frequency $\omega_d$ and phase $\phi_d$. For long gate times, $\omega_d$ is close to the qubit frequency, but taking it as a variable parameter allows the optimizer to account for small AC stark shifts and results in smoother pulse shapes for the remaining envelope. For the two-qubit gate there is no resonance condition and therefore no carrier is included. 

For the optimization of filtered control pulses discussed in Sec. \ref{sec:bw} we must take into account that the ideal control signal is shifted compared to the actual driving signal. In this case we replace $\Omega_c(t)\rightarrow \Omega_c(t+t_d)$ and $\Phi_c(t)\rightarrow \Phi_c(t+t_d)$, which are non-zero in the interval $[-t_d, t_g-t_d]$. Otherwise the ansatz for the control pulses is left unchanged. The value of $t_d$ is determined at each iteration of the optimization algorithm by finding the peak position of the cross-correlation function between the input and the filtered pulse.

\subsection{Cost function}
For the optimization we must choose a cost function $\mathcal{C}$, which for unfiltered pulses and for single-qubit rotations we take as the gate error $\mathcal{C}=\mathcal{E}=1-\mathcal{F}$, where the fidelity $\mathcal{F}$ is defined in Eq. (\ref{eq:fidelity}). For the two-qubit gate we only require that the exact unitary has the same form as $V$ defined in Eq. (\ref{eq:V}), independently of  the precise values of $\beta$ and $\gamma$. Therefore, we choose the cost function $\mathcal{C}_V$ given in Eq. (\ref{eq:cost_2q}).

For the optimization of filtered control pulses we want to minimize the support of the driving signals outside the target interval $[0,t_g]$. Therefore, for the single-qubit gates we consider the generalized cost function 
\begin{gather}
	\mathcal{C} = \mathcal{E} +\eta\dfrac{\zeta(-t_d,0)+\zeta(t_g,10 t_g)}{\zeta(0,t_g)},\label{eq:1q_penalty}
\end{gather}
where
\begin{gather}
	 \zeta(t_i,t_f) = \int_{{t_i}}^{{t_f}} {|\Omega(t+t_d)|} \: d{t}.
\end{gather}
Similarly, for the two-qubit gate we use
\begin{gather}
	\mathcal{C} = \mathcal{C}_V +\eta\dfrac{\xi(-t_d,0)+\xi(t_g,10 t_g)}{\xi(0,t_g)},\label{eq:2q_penalty}
\end{gather}
where 
\begin{gather}
 \xi(t_i,t_f) = \int_{{t_i}}^{{t_f}} {\left|\Phi_c(t+t_d)\right|} \: d{t}.
\end{gather}
In both cases $\eta$ is an additional penalty coefficient.  For the current problem we find that a range of $\eta\in [0.01,0.1]$ is a suitable choice to obtain very good optimization results in all parameter regimes. We take $\eta=0.01,~0.1$ for Fig. \ref{fig:8_Filter}(c) and Fig. \ref{fig:8_Filter}(d), respectively.

\subsection{Initial conditions and pre-optimization}
The performance and convergence of numerical optimizers depend strongly on the choice of the initial conditions for the pulse parameters. For all the single-qubit gates $R_{n(\phi)}(\theta)$ we start the optimization with the parameters $\omega_{d}=\omega_{10}$, $\phi_d=-\phi$, $a_1=1$ and $a_{n>1}=0$.  
For the unfiltered pulses we then set the initial value of the driving strength to $\Omega_0=\pi \theta/(2t_g\varphi_{10})$, which would implement the correct rotation under the validity of the rotating-wave and the two-level approximation. For the filtered pulses we estimate the reduction of the pulse area of the filtered pulse within the time window $[0,t_g]$ and adjust the initial value of $\Omega_0$ accordingly.  For the two-qubit gates we initialize the control pulse with a slightly lower amplitude, $a_1=\pi/2$ and $a_{n>1}=0$, such that the interaction still scales approximately linearly in $\Phi_c(t)$.  
The initial value for the Josephson energy is set to $E_{J_s}/\hbar = \pi/(4W)$, where
\begin{equation}
	W =\int_{t_d}^{t_g+t_d} \textrm{d}t \int_{-\infty}^{t}\textrm{d}t' \,  F(t-t') \sin\left[\frac{a_1}{2}\sin\left(\frac{\pi t'}{t_g}\right)\right]
\end{equation}
for both the unfiltered ($t_d=0$) and the filtered ($t_d>0$) pulses.

For very short gate times, the initialization of parameters discussed above still gives a rather poor approximation for the optimal pulses. Therefore, as a second step we implement an additional pre-optimization step for the first $n=1,\dots, 5$ amplitudes $a_n$ and the frequency and the phase of the carrier. For this pre-optimization we use the Nelder-Mead algorithm \cite{Nelder1965} with a reduced cost function $\mathcal{C}_{\textrm{pre}} = \mathcal{C}/10$. This gradient-free algorithm is very efficient for a small parameter space and thus allows us to substantially improve the initial conditions without significantly increasing the computation time. We set the target infidelity of the optimization to $10^{-5}$, but run at most 2000 iterations. A similar hybrid optimization scheme combining gradient-free and gradient-based methods has been shown to outperform the schemes with either one of the two methods alone \cite{Goerz2015}.

\subsection{Gradient-based optimization}\label{sub:gradient_based_optimization}
After obtaining good initial values we use the gradient-based limited-memory Broyden–Fletcher–Goldfarb–Shanno (L-BFGS) algorithm \cite{Nocedal1980} for the actual optimization. At each iteration step this algorithm requires the evaluation of the cost function $\mathcal{C}$ as well as its gradient with respect to all variable parameters. For problems with a small Hilbert space dimension, the evaluation of this gradient can be done using, for example, the GOAT algorithm described in Ref.~\cite{Machnes2018}. However, for the current problem it turns out to be more efficient to calculate $\mathcal{C}$ multiple times and approximate its gradient by a finite difference method. To keep the pulse spectrum simple, we divide the set of $\{a_n\}$ into two batches, i.e., $\{a_1,...,a_{10}\}$ and $\{a_{11},...,a_{20}\}$, and optimize them alternatively together with the carrier parameters. We use 15 optimization steps for each batch and repeat the sequence five times. Similar to the pre-optimization step, we stop the search already before in case an infidelity below $10^{-5}$ has been reached. All our numerical simulations are implemented in \emph{Julia} using the packages \emph{Optim.jl} \cite{Mogensen2018} for optimization, \emph{OrdinaryDiffEqs.jl} \cite{Rackauckas2017} for evaluating the time-evolution operator, \emph{DSP.jl} for filtering and \emph{QuantumOptics.jl} \cite{Kramer2018} for constructing operators and simulating the composite circuit results.

\begin{figure}[t]
	\centering
	\includegraphics[width=\linewidth]{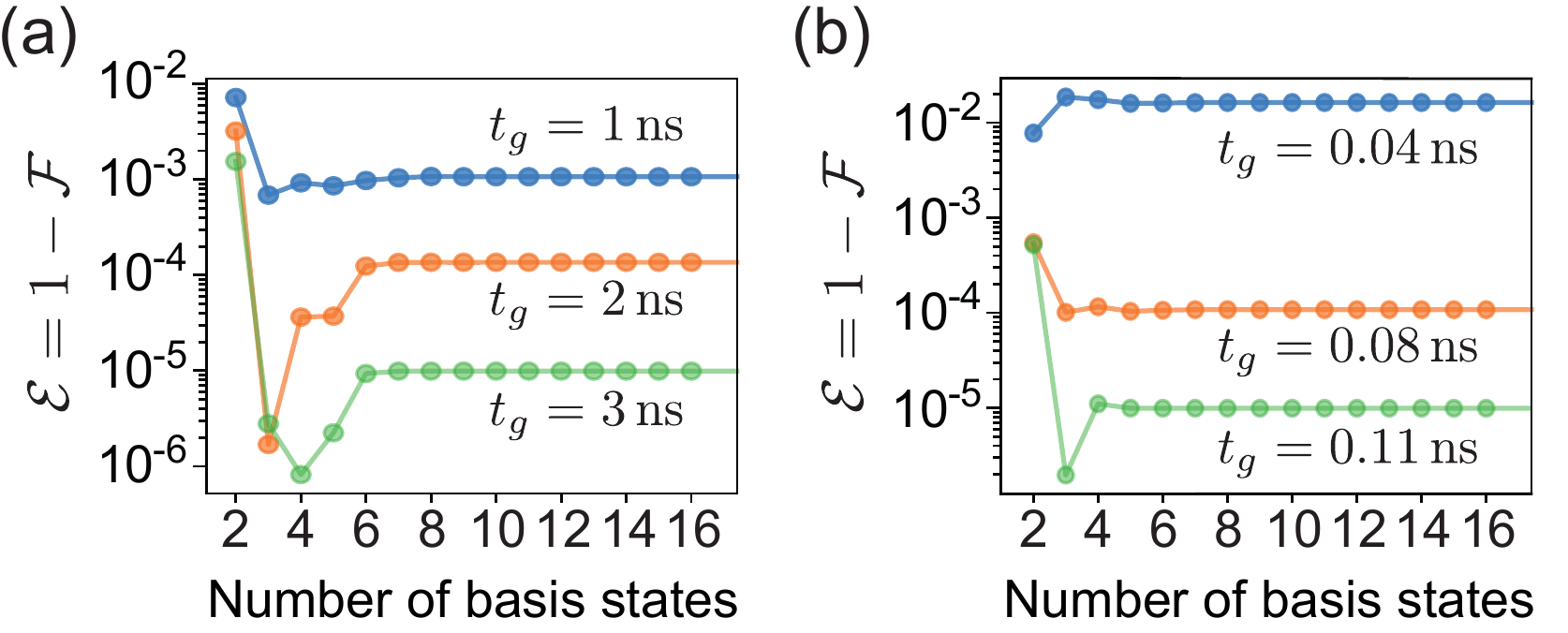}
	\caption{Level truncation. Plot of the gate error $\mathcal{E}$ for an optimized $R_Y(\pi/2)$ rotation  as a function of the number of included basis states and for different gate times. The plot in (a) shows the case of a transmon qubit with a negative anharmonicity of $\alpha_r\simeq-0.2$ and (b) the results for a qubit with a positive anharmonicity of $\alpha_r\simeq 5.0$. }%
	\label{fig:full-model}
\end{figure}

\subsection{Level truncation}
In all our numerical simulations we truncate the Hilbert space of each subcircuit to the ten lowest basis states. To show that this number of basis states is sufficient, we plot in Fig.~\ref{fig:full-model} the results for the gate error $\mathcal{E}$ as a function of the number of included basis states for the two examples of (a) a  transmon qubit with a small negative anharmonicity and (b) a flux qubit with a large positive anharmonicity. We see that for systems with a negative anharmonicity, there can be accidental multi-photon resonances with higher states and we need at least six levels to get convergent results. For systems with positive anharmonicity the inclusion of four to five levels is usually enough and the same is true for systems with small positive nonlinearities. Note that in both examples the results obtained from a three-level truncation, as often assumed in optimization studies, are still inaccurate.

\section{Elimination of $Z$ rotations}\label{app:ZRotations}
Although  in Sec.~\ref{subsec:Zrotations} we have shown that rather fast $Z$ rotations can be realized without changing the control Hamiltonian, in many circuits the execution of $R_Z$ gates can be avoided altogether by constructing equivalent circuits according to the following rules: 

\begin{enumerate}
    \item By making use of the commutation relation $R_Z(\theta) R_{n(\phi)}(\varphi) = R_{n(\phi + \theta)}(\varphi) R_Z(\theta)$ the $ Z $ rotations can be exchanged with preceding or successive $R_{n(\phi)}$ gates.  All $R_Z$ gates that can be moved to the beginning or the end of the circuit through this procedure can be dropped, assuming that we initialize and measure the qubits in the computational basis.

  \item The $ Z $ rotations that are initially located between two $ V $ gates are commuted either right next to the following  or the preceding $ V $ gate. 
    We can then use 
    \begin{align}
     &\left[R_Z(\varphi) \otimes R_Z(\theta)\right]V \nonumber\\
     &=\left[I\otimes R_Z(\theta - \varphi) \right]V \left[R_Z(\varphi)\otimes R_Z(\varphi)\right],
    \end{align}
    to bring the common part of the rotation to the other side of the two-qubit gate and proceed with step 1. 
    
   \item After implementing steps 1 and 2 we are left with a circuit where at most one $Z$ rotation appears between two $V$ gates. These remaining gates are decomposed as $R_Z(\theta)= R_{n(\phi)}(\pi/2)R_{n(\phi)+\pi/2}(\theta)R_{n(\phi)}(-\pi/2)$ and combined with neighboring $R_{n(\phi)}$ gates.     

\end{enumerate}  
Since the duration of an $R_{n(\phi)}$ gate depends on the rotation axis $n(\phi)$, step 1 changes the duration of the remaining single-qubit gates, but on average this effect cancels out. In step 3 the rotation axis $n(\phi)$ can be optimized and depending on the neighboring gates the maximum added gate count per $ Z $ rotation is two. This only happens in the unlikely situation when an $ R_Z $ rotation is sandwiched between two $ V $ gates with no neighbouring single-qubit rotations, i.e., $ V[I\otimes R_Z(\varphi)]V $. Therefore,  for most applications  $R_Z$ rotations can be eliminated or replaced by equivalent $R_{n(\phi)}$ with only a minor overhead in the total computation time.

\section{Effective nonlinearities during two-qubit gates}\label{app:Outliers} 
In the optimization of the fidelities for the two-qubit gate shown in Fig.~\ref{fig:7_V_optimization}(a), we find isolated points where despite a rather small qubit nonlinearity parameter $\alpha_r$, rather fast gates can be implemented, which break the overall bound set by $\pi/\alpha$. The existence of such outliers can be understood from the fact that in the two-qubit case the coupling junction can induce additional nonlinearities, which are not taken into account in the definition of the single-qubit nonlinearity $\alpha_r$. 

To illustrate this point in more detail, we consider in this appendix the Hamiltonian
\begin{equation}
H_{\rm eff}(t)= H_q^{(1)}+ E_{J_s}\cos\left(\frac{\Phi_s(t)}{2\Phi_0}\right)\cos\left( \varphi_1 \right),
\end{equation}  
which represents the coupling circuit shown in Fig.~\ref{fig:2_Interactions}(a), but with the phase of the second qubit set to $\varphi_2=0$. For this effective single-qubit circuit we can define an instantaneous nonlinearity parameter,
\begin{equation}\label{eq:alphaeff}
	\alpha_{\rm eff}(t)= (E_2(t)-E_1(t))-(E_1(t)-E_0(t)),
\end{equation}
where the $E_i(t)$ are the instantaneous eigenenergies of $H_{\rm eff}(t)$. This parameter reflects the degree of single-qubit nonlinearity at each point in time during the two-qubit gate sequence.

\begin{figure}[t]
	\centering
	\includegraphics[width=\linewidth]{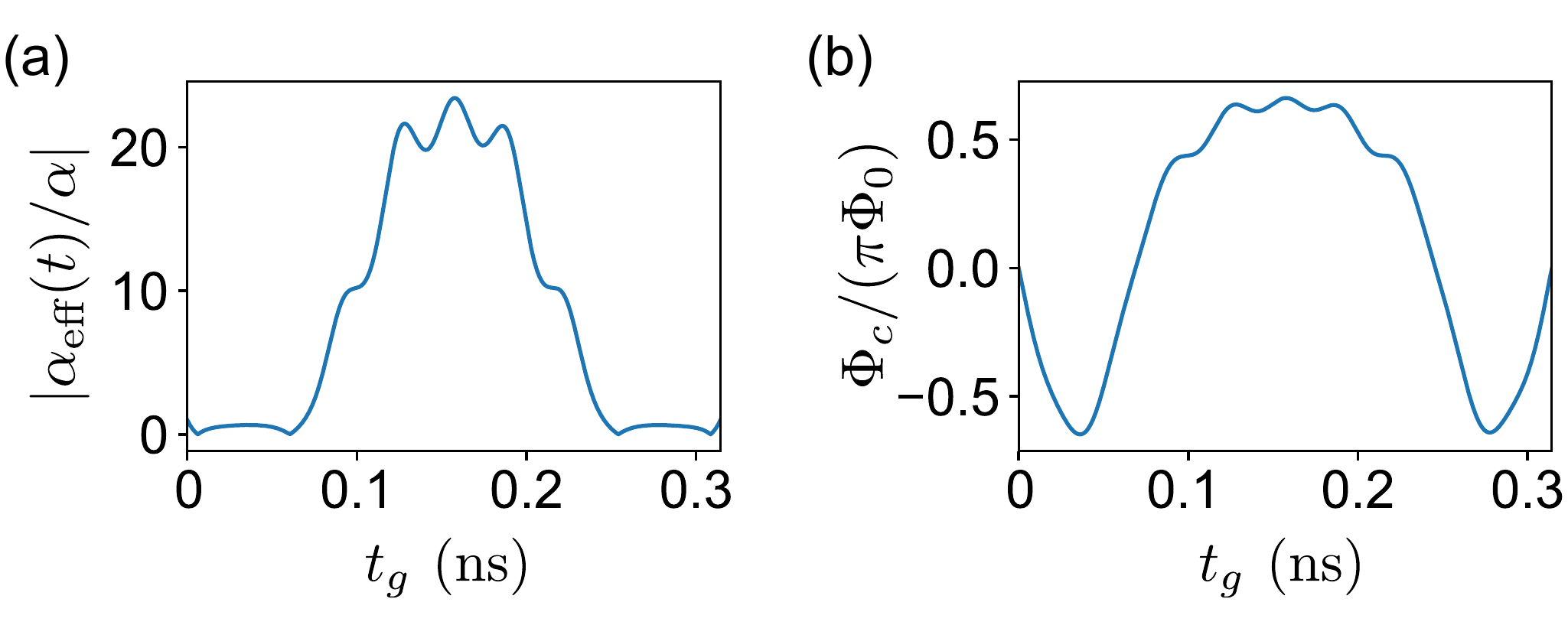}
	\caption{(a) Plot of the effective nonlinearity parameter defined in Eq.~\eqref{eq:alphaeff} for the control pulse shown in (b). This optimized control pulse has been obtained for $\alpha_r=0.05$ and implements the two-qubit $V$ gate in a time of $t_g\simeq 300$ ps with a gate error of $\mathcal{E}\simeq 0.004$.}
	\label{fig:nonlinearity}
\end{figure}

In Fig.~\ref{fig:nonlinearity}(a) we plot this parameter for the case $\alpha_r=0.05$ and $t_g\simeq 300$ ps, where it is possible to implement a high-fidelity two-qubit gate in a time much faster than $\pi/\alpha\simeq 2$ ns. We see that in this example the effective nonlinearity during the pulse is considerably larger than the bare nonlinearity, with a time-averaged value of about $\bar \alpha_{\rm eff}\simeq 8.1 \alpha$. This explains the existence of such outliers, but also shows that the achievable gate time still respects the bound $t_g\geq \pi/\bar{\alpha}_{\rm eff}$, when expressed in terms of the effective nonlinearity. 
However, since the occurrence of such exceptional pulses is rather rare and they are no longer observed when introducing a realistic filter-bandwidth, we don't go into further details here.

\end{document}